\documentclass[a4paper,12pt]{article}
\usepackage[left=2.5cm,right=2.5cm,top=1cm, bottom=2cm]{geometry}
\usepackage{multirow}
\usepackage{colortbl}
\usepackage{color}
\usepackage[pdftex]{graphicx}
\usepackage{mathtools}
\usepackage{amsmath}
\usepackage{amssymb}
\usepackage{url}

\usepackage[table]{xcolor}
\usepackage{authblk}

\global\long\def\dt#1{\frac{\mathrm{d}#1}{\mathrm{d}t}}

\begin{document}

\title{ The effect of intermittent upwelling events on plankton blooms.}

\author{Ksenia Guseva and Ulrike Feudel}

\affil{Theoretical Physics/Complex Systems, ICM, University of Oldenburg, 26129 Oldenburg, Germany}

\date{}
\maketitle

\begin{abstract}
  In the marine environment biological processes are strongly affected by
  oceanic currents, particularly by eddies (vortices) formed by the hydrodynamic
  flow field. Employing a kinematic flow field coupled to a population dynamical
  model for plankton growth, we study the impact of an intermittent upwelling of
  nutrients on triggering harmful algal blooms (HABs). Though it is widely
  believed that additional nutrients boost the formation of HABs or algal blooms
  in general, we show that the response of the plankton to nutrient plumes
  depends crucially on the mesoscale hydrodynamic flow structure. In general,
  nutrients can either be quickly washed out from the observation area, or can
  be captured by the vortices in the flow. The occurrence of either scenario
  depends on the relation between the time scales of the vortex formation and
  nutrient upwelling as well as the time instants at which upwelling pulse
  occurs and how long do they last. We show that these two scenarios result in
  very different responses in plankton dynamics which makes it very difficult to
  predict, whether nutrient upwelling will lead to a HAB or not. This explains,
  why observational data are sometimes inconclusive establishing a correlation
  between upwelling events and plankton blooms.

\end{abstract}


\section{Introduction} 

Coastal regions susceptible to harmful algal bloom (HAB) events are often
subjected to upwelling \cite{pettersson_monitoring_2013}. Due to this upwelling
nutrient-rich deep waters are transported into the euphotic zone and this inflow
fosters favorable conditions for the growth of algae
\cite{mann_dynamics_2005}. As recent studies notice a significant increase of
the number of harmful algal bloom events in the whole world
\cite{belgrano_north_1999, sellner_harmful_2003,kahru_ocean_2008}, it becomes
imperative to understand the interplay between the biotic and physical factors
that work as their trigger.

Lateral mixing and stirring by the hydrodynamic flow redistributes the nutrients
and the suspended microorganisms, shaping the spatial heterogeneity of the
marine ecosystem at different scales~\cite{martin_phytoplankton_2003}, leading
to plankton blooms, which exhibit a non-uniform distribution in space, referred
to as ``patchines''. This non-uniformity was ubiquitously detected around the
globe by satellite imagery~\cite{gower_phytoplankton_1980,
  mcgillicuddy_mechanisms_2016, levy_modulation_2008,
  lehahn_satellite-based_2018} and by samples along ship transects
\cite{mackas_spectral_1979, martin_plankton_2002, weber_variance_1986}. In a
seminal work by Abraham~\cite{abraham_generation_1998}, a very simple model of
turbulent transport was able to reproduce this spatial heterogeneity in the
plankton distribution and its statistical properties, such as spectra. This
model shows that advection by ocean flows on the mesoscale (10 - 100 km) can
spatially distort the concentration of plankton leading to the development of
small spatial patterns and thin filaments. Subsequent theoretical studies have
observed that the ratio between biological and hydrodynamic flow time scales has
a non-trivial impact on how plankton is distributed
spatially~\cite{bracco_horizontal_2009, mckiver_influence_2009}. The flow field
influences not only the spatial distribution but also the abundance of
plankton. Further studies have shown that it is possible to trigger or suppress
HABs by tuning the flow to the biological
timescales~\cite{mckiver_plankton_2009,
  hernandez-garcia_sustained_2004}. Therefore the hydrodynamics plays a central
role for phytoplankton ecosystems, not only with respect to its spatial patterns
but also to the inter-specific interactions, establishing so called ``fluid
dynamical niches'', which provide particular growth conditions for certain
species~\cite{dovidio_fluid_2010, vortmeyer-kley_eddies:_2019}. Coherent
structures of the flow field, such as, for instance eddies play an important
role influencing the biological processes in the ocean. The recent advances in
detecting and tracking eddies in the ocean have shown that they often are long
lived. Notably they can trap fluid and the whole community of plankton and
bacteria inside, which affects the diversity and dominance structure of
phytoplankton species observed in the system~\cite{levy_dynamical_2015,
  vortmeyer-kley_eddies:_2019}. On the one hand, the species in the almost
isolated ecosystem inside the eddy are subjected to competitive pressure.  On
the other hand, theoretical models speculate that due to this trapping the
organisms can also be sheltered inside the eddy from predators or competitors, a
mechanism proposed as a possible explanation for the coexistence of
species~\cite{karolyi_chaotic_2000, scheuring_competing_2003, tel_chemical_2005,
  bracco_mesoscale_2000}. Both effects are a direct result of transport barriers
established by the flow field.

The productivity enhancing effect of coastal upwelling is also shown to be
strongly affected by the presence of coherent structures in the flow
field~\cite{rossi_comparative_2008,rossi_surface_2009,
  gruber_eddy-induced_2011}. The eddies mix and disperse nutrients, while also
taking them away from the coastal region.  This leads to a decrease in the
primary production near the shore, as was recorded for eastern boundary
upwelling systems~\cite{gruber_eddy-induced_2011}. Furthermore, these nutrients
while being transported offshore by the eddies may also trigger the growth of
the associated phytoplankton. Theoretical works have shown that eddies in this
case work as incubators for growth by sustaining favorable environmental
conditions. These models emphasize the importance of the role of biological and
hydrodynamic timescales in triggering plankton blooms and specifically HABs in
this scenario~\cite{sandulescu_kinematic_2006, sandulescu_plankton_2007,
  sandulescu_biological_2008, bastine_inhomogeneous_2010}. However, these
studies, have so far only analysed the conditions of an upwelling which is
constant in time. Nevertheless, upwelling itself is not a steady process since
it depends on winds and seasonality, being therefore highly
intermittent. Furthermore, from observations of HABs in nature, it is not always
possible to correlate the strength and the duration of an upwelling event and
the occurrence and magnitude of HABs. While it was possible to establish such a
direct relation for some species (e.g. diatoms of genus {\it
  Pseudo-nitzschia}~\cite{trainer_domoic_2000,kudela_harmful_2005} and some
dinoflagelates ~\cite{omand_episodic_2012}), for others a more complex chain of
events appears to be driving the
outcome~\cite{bialonski_phytoplankton_2016}. Moreover, the major challenge
consists in finding out how the occurrences of HABs and the episodic upwelling
events are associated on a local scale~\cite{bialonski_phytoplankton_2016,
  nezlin_phytoplankton_2012}.

In this work we analyse the impact of intermittent upwelling events on
phytoplankton growth and changes in dominance patterns in the presence of
mesoscale hydrodynamic structures for a biological system with three trophic
levels. We modify the reaction-advection-diffusion model introduced by
Sandulescu et al. (2006) for the area around the Canary islands
~\cite{sandulescu_kinematic_2006}, that couples advection by a vortex street
behind an island with a model of plankton dynamics. As in
~\cite{sandulescu_kinematic_2006} we also choose to ignore the possibility of
eddy-induced Ekman pumping, a well known phenomenon where circulating ocean
currents bring nutrients upwards or downwards within the eddies
~\cite{martin_mechanisms_2001,gaube_satellite_2013}. In this way we can isolate
the plankton's response to a single upwelling region, and study the effects of
upwelling intermittency. Furthermore it allows us to simplify to horizontal
advection only. In contrast to ~\cite{sandulescu_kinematic_2006}, we analyse
here a community that consists of two phytoplankton species competing for a
limiting resource and grazed by zooplankton. The population model
~\cite{chakraborty_harmful_2014} chosen displays excitability, which arises from
the interplay of the fast dynamic timescale of phytoplankton growth (activator)
with a slow development of zooplankton (the inhibitor). We chose two scenarios
with different plankton communities: (I) where the community structure is shaped
only by the availability of nutrients in the environment; (II) where both, the
grazing pressure and the nutrient availability, trigger the bloom
formation. First we show that these two systems exhibit very different
spatio-temporal dominance patterns and display distinct and characteristic
dynamical reactions to an upwelling event. Then we show that for both
parameterizations even identical pulses of nutrient influx, trigger a diverse
set of reactions in the plankton dynamics. The variety of possible responses can
only be understood by analysing the interplay of different time scales that
characterise the system as well as the interplay between the upwelling and
mesoscale hydrodynamic structures present in the flow. The outcome is even more
complex for the case of irregular pulses with a variety of strengths and
durations. Our analysis shows that it is impossible to establish a relation
between the HABs formation and upwelling events, by only looking at the
respective time series of nutrients and plankton abundances without considering
the mesoscale mixing by the ocean flow in the observation region.

The work is organised as follows: First, in Sec.~\ref{sec:model} we briefly
introduce the coupled hydrodynamic-biological model used. In
Sec.~\ref{sec:exit}, we examine how the position and initial time of nutrient
parcels initialized at the upwelling region affect the residence time of
nutrients in the observed area. In Sec.~\ref{sec:spacetime} we describe the
spatio-temporal patterns of plankton for the scenario without upwelling and for
the scenario subjected to a single upwelling pulse. Next, in
Sec.~\ref{sec:init} we describe how the response of the populations varies
considering different initial times for an upwelling pulse. Furthermore, we
analyse the chance of HAB formation for upwelling pulses of different duration
and strengths. In Sec.~\ref{sec:inter} we extend our analysis to the study of
the response to a series of irregular (intermittent) upwelling events. We
conclude in Sec.~\ref{sec:conclusion}.

\section{Model}\label{sec:model}

\begin{figure}[h!]
\centering
\includegraphics[scale=.7]{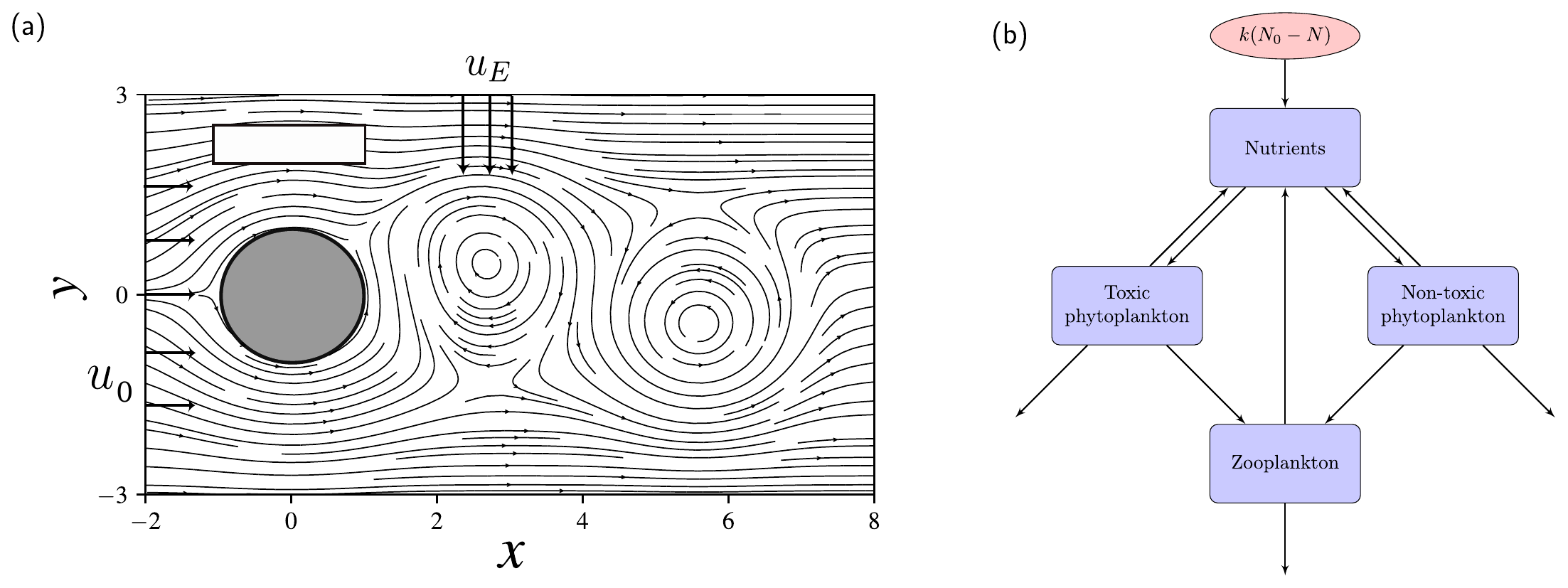}
\caption{(a) Two dimensional flow field of the vortex street behind an
  island. The white rectangular area above the island (gray cylinder) sketches
  an upwelling region, while the arrows symbolise the Ekman flow $u_E$
  perpendicular to the main flow $u_0$ (b) Schematic representation of the
  biological model.}\label{fig:model}
\end{figure}

This section describes the modeling framework used in this work. The model
consists of a two dimensional kinematic velocity field coupled to a biological
model, see Fig.~\ref{fig:model}.

\subsection{Hydrodynamic model}

The hydrodynamic model is represented by a two dimensional kinematic velocity
field $(u_x, u_y)$ that flows trough a predefined observation region passing by
a circular obstacle (of radius $r$), located at $(x_0, y_0) = (0,0)$, mimicking
an island. The flow velocity is such that it allows for the formation of
vortices in the wake of the island. These vortices are released and carried away
from the island along the observation region, from left to right in
Fig.\ref{fig:model} (a). Although this flow field can be obtained as a solution
of the Navier-Stokes equation, we use an analytically generated
field~\cite{jung_application_1993} that captures the main characteristics of
this solution, but with smaller numerical efforts. The flow is characterized by
a period T and in this approach a predefined stream function $\Phi$ is used to
generate it: $u_x = \frac{\partial \Phi}{\partial y}$,
$u_y = -\frac{\partial \Phi}{\partial x}$. This flow is known as an open chaotic
flow in literature, for a detailed description of the modeling approach and the
stream function see ~\cite{jung_application_1993, sandulescu_kinematic_2006}.
The model has been parameterized to represent one of the islands of the
Canarian Archipelago, located in the Eastern Boundary Upwelling System off the
African Coast, in agreement with~\cite{sandulescu_kinematic_2006}, for details
see Supplemental material.

\subsection{Biological model}\label{sec:bio_model}
\begin{figure}[h!]
  \includegraphics[scale=.9]{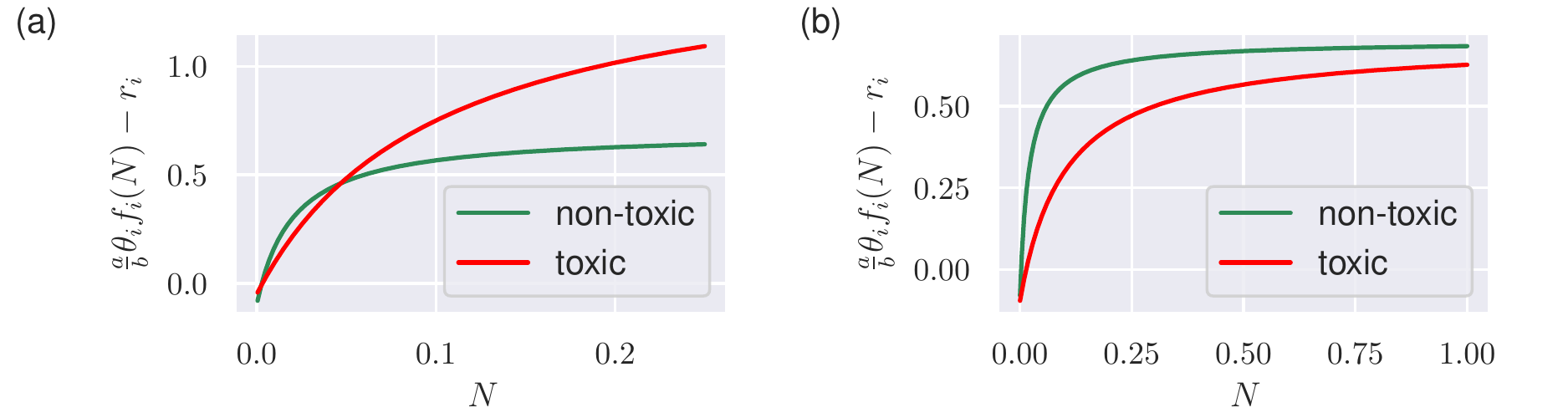}
  \includegraphics[scale=0.7]{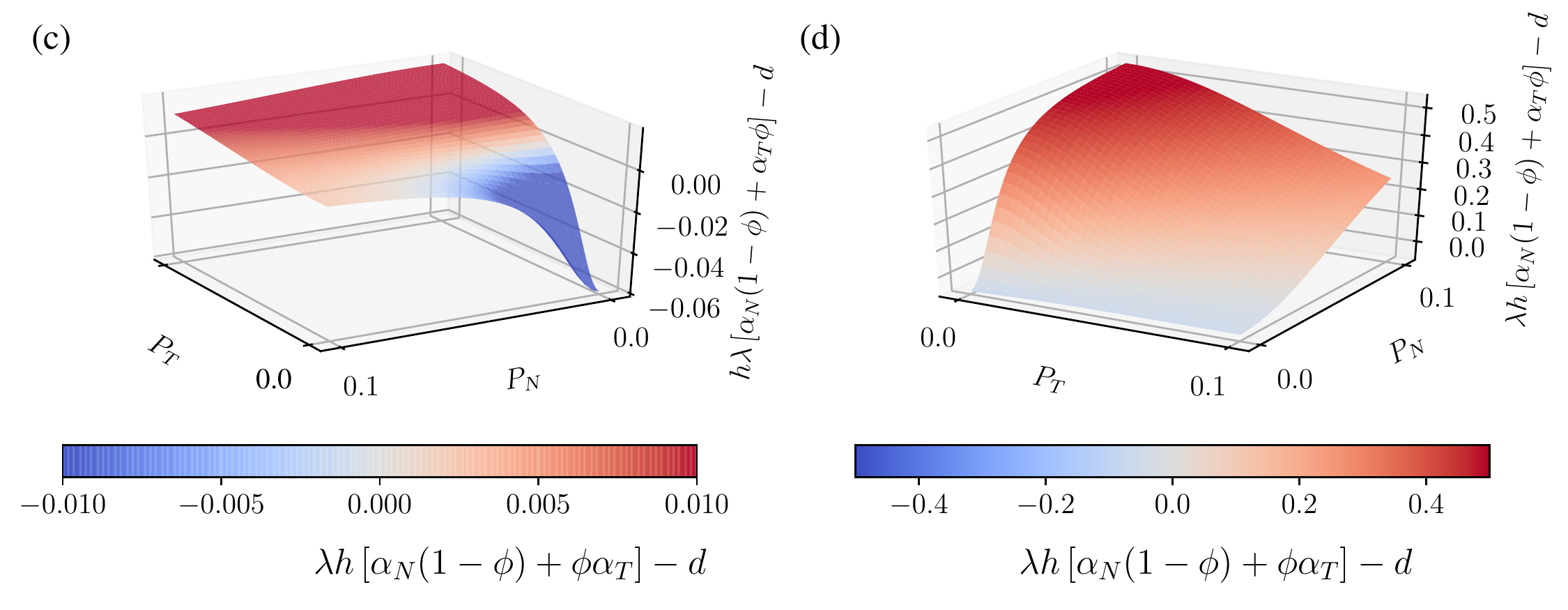}
  \centering
  \caption{(a, b) Approximate net-growth rate of the two phytoplankton groups
    (it neglects the self- shading function and the grazing by zooplankton) for
    System (I) and System (II) respectively. (c, d) Approximate net-growth rate
    of the zooplankton feeding as a function of concentrations of toxic
    and non-toxic species of phytoplankton for: (c) system (I) and (d)
    system (II).}\label{fig:systems}
\end{figure}

The biological model used, consists of a food web with three trophic levels NPPZ
(Nutrients, two Phytoplankton species and Zooplankton) formulated in
~\cite{chakraborty_harmful_2014} to describe the formation of harmful algal
blooms (HABs). One of the phytoplankton species is considered to be toxic and the
other one non-toxic, their concentrations are $P_T$ and $P_N$,
respectively. They compete for a limiting nutrient resource, $N$, while being
grazed by zooplankton, $Z$ (see Fig.~\ref{fig:model} (b)). The inter- and
intraspecific interactions are described by:
\begin{align}\label{eq:pop_model}
  \dt N &= k[N_0-N] - g(P_N,P_T)\left[f_N(N) P_N+ f_T(N)P_T \right] + r_NP_N+r_TP_T\nonumber\\
        &+\beta h(P_N, P_T)\left[\lambda(1-\phi)P_N^2 + \lambda \phi P_T^2\right]Z + \gamma dZ,\nonumber\\
  \dt{P_N} &= \theta_Nf_N(N)g(P_N, P_T)P_N - r_NP_N - \lambda[1-\phi]h(P_N, P_T)P_N^2Z -sP_N,\\
  \dt {P_T} &= \theta_Tf_T(N)g(P_N, P_T)P_T - r_TP_T - \lambda\phi h(P_N, P_T) P_T^2 Z -sP_T,\nonumber\\
  \dt Z &= \left[\alpha_N\lambda(1-\phi) P_N^2 + \alpha_T \lambda\phi P_T^2\right]h(P_N, P_T) Z -dZ.\nonumber
\end{align}
The functional responses and parameterization are listed in the Supplemental
material. As we will show below the system's response to nutrient influx from
upwelling results from a combination of bottom-up and top-down controls. To be
able to analyse how these controls drive HAB formation we chose two different
parameterizations for the population model: system (I), where the community
structure and the dominant species results mainly from the availability of
nutrients in the environment; system (II), where the grazing preference of
zooplankton together with the nutrient availability both establish the resulting
community structure. The parameters chosen for the two systems are very similar,
with a few differences which emphasize different ecological processes. In system
(I) the nutrient conversion rate $\theta_i$ and the respiration rates $r_i$ are
different for each species ($\theta_N < \theta_T, r_N > r_T$), so that the net
growth rate of $P_T$ is always larger than the one of $P_N$ (see
Fig.~\ref{fig:systems}(a)). On the contrary, in system (II) these parameters are
set to the same values ($\theta_N = \theta_T, r_N = r_T$), for both
species. Consequently, in system (II) the abundance of nutrients cannot drive a
dominance change (see Fig.~\ref{fig:systems}(b)). To especially test the role of
grazing, we modify the parametrisation of zooplankton in the second system, in
order to have a stronger influence of the grazer within the food chain. To that
end, we boost the abundance of zooplankton by increasing its maximum grazing
rate $\lambda$ and its growth efficiency on the non-toxic species
$\alpha_N$. Finally we also add in this set up a strong preference of
zooplankton to feed on non-toxic species ($\phi = 0.05$), contrarily to system
(I) where there is no preference ($\phi = 0.5$). With this change, system (II)
has a very strong top down control by design. This is demonstrated in
Fig.~\ref{fig:systems} (c, d) that show an approximate net growth of zooplankton
for the two systems. While for system (II) the net growth rate of zooplankton is
always positive Fig.~\ref{fig:systems} (d), this does not hold for system
(I). For very low abundances of phytoplankton, there is not enough food for
zooplankton to survive. Therefore, for very low nutrient supply and subsequently
very low phytoplankton abundance, zooplankton would go extinct.

As described previously the model takes into account the vertical influx of
nutrients from the deep ocean into the mixed layer where all biological
processes take place.  The rate of this influx is given by $k$.  This influx of
nutrients may occur due to turbulent diffusion ($k = k_d$) or by vertical
transport due to upwelling ($k = k_{\text{up}}$).  The diffusive flux is
$D_v \frac{\partial^2 N }{\partial z^2} \sim D_v\frac{(N_0 - N)}{h^2} $, where
$N_0$ is the concentration of nutrients below the mixed layer and $h$ is an
average extension of the gradient. By using the definition
$k_d = \frac{D_v}{h^2}$ we can rewrite the relation as $k_d(N_0 - N)$. For the
ocean we find in the literature values of $D_v \sim 0.1 - 2.6$ m$^2$
day$^{-1}$~\cite{mann_dynamics_2005, martin_patchy_2002}. We use the known
extension of the thermocline to estimate $h$ and therefore adopt values from
$10$ to $25$ m. With these parameters we can evaluate $k_d$ in the range of
$10^{-2}$ --- $10^{-4}$ day$^{-1}$. The nutrient transport due to upwelling, on
the other hand, is defined as
$u_z \frac{\partial N }{\partial z} \sim u_z \frac{\Delta N}{\Delta z}$.
Therefore for the situation with upwelling we can define the thermocline
exchange rate as $k_{\text{up}}(N_0 -N)$ with $k_{\text{up}} = \frac{u_z}{h}$ +
$k_d$. It is known that the vertical velocity $u_z$ may reach values as large as
$\sim 40\;$ m day$^{-1}$~\cite{martin_patchy_2002}, however specifically for the
region of the Canarian Archipelago we find estimations close to $10\;$ m
day$^{-1}$~\cite{barton_canary_2001}. This gives us $k_{\text{up}} $ of 1
day$^{-1}$. Therefore in this work we restrict ourselves to value for
$k_{\text{up}}$ of the order of unity.

\begin{figure}[h!]
\includegraphics[scale=.75]{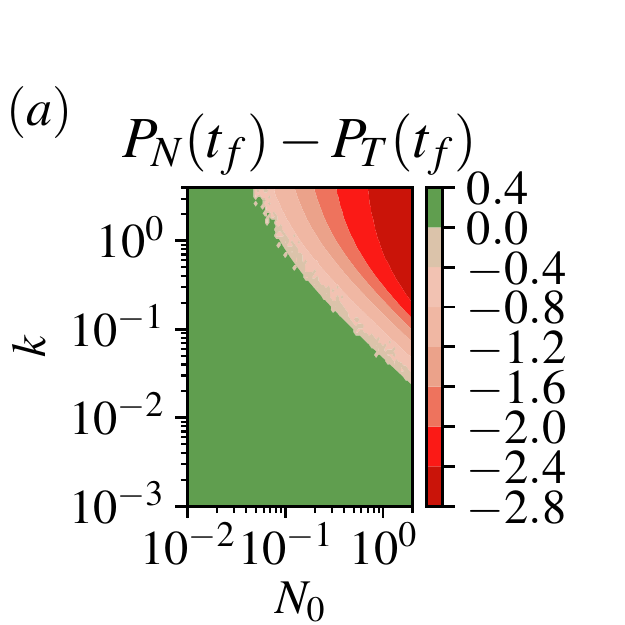}
\includegraphics[scale=.75]{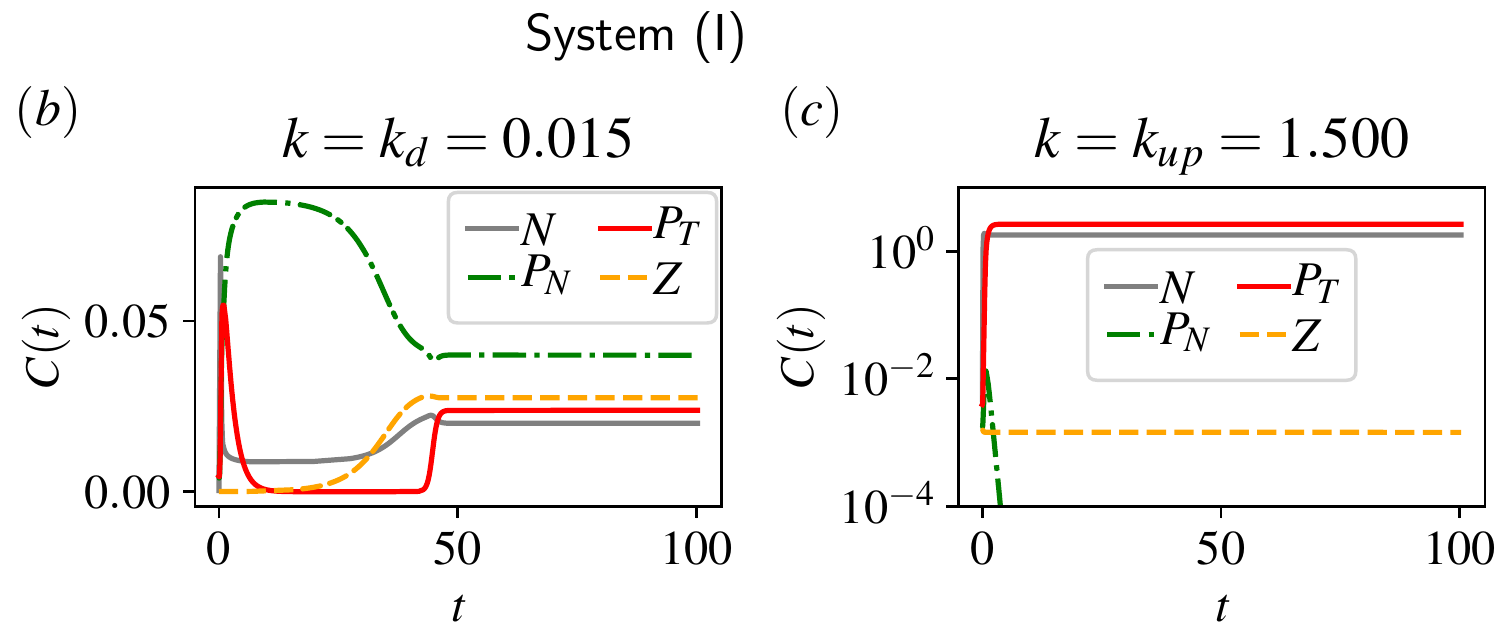}
\includegraphics[scale=.75]{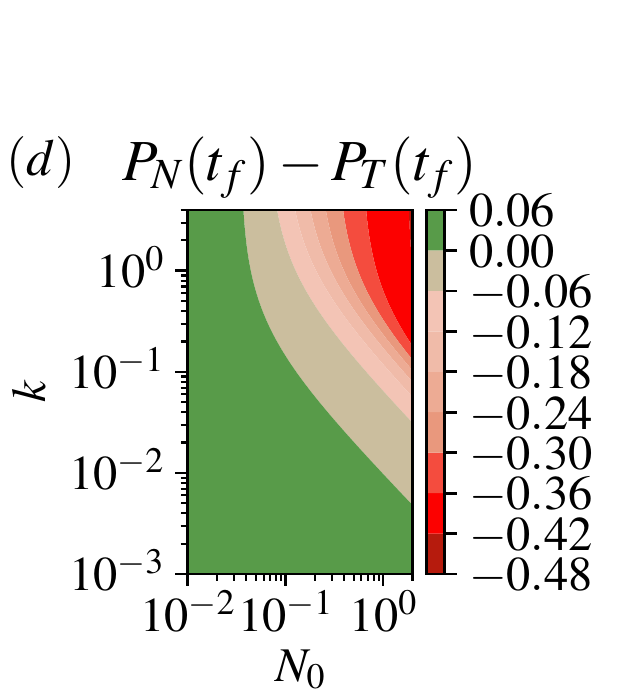}
\includegraphics[scale=.75]{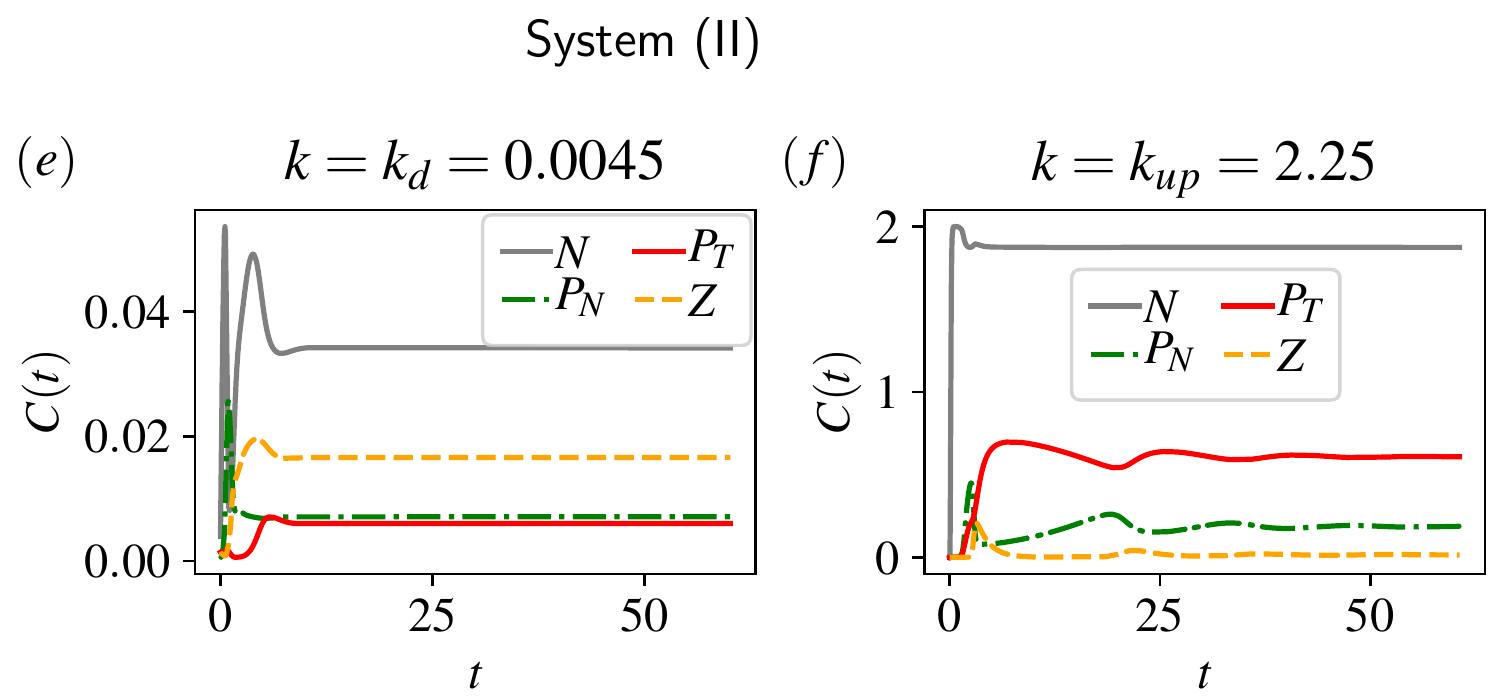}
\caption{(a,d) Difference between the concentrations of the non-toxic $P_N$ and
  toxic $P_T$ species for a range of values of the thermocline exchange rate $k$
  and of concentrations of nutrients below the thermocline; (b, c, e, f) time
  evolution of the biological model for low (b, e) and high (c, f) cross thermocline
  exchange rates $k$.}\label{fig:system_one}
\end{figure}

As already mentioned the nutrient influx is regulated by two parameters: the
cross thermocline exchange rate $k$ and the nutrient concentration below the
thermocline $N_0$. It is therefore compelling to outline here how the coupled
effect of these parameters is reflected in the steady state community
structure. These results are summarised in Fig.~\ref{fig:system_one} (a) and (d)
for system (I) and (II) respectively. The region of dominance of the toxic
species is shown in red, and of the non-toxic in green.  We also set-up two
values for the cross thermocline exchange: $k_d$ and $k_{\text{up}}$
representing the conditions without and with upwelling respectively.
  
Our choice for system (I) corresponds to $k_d = 0.015$ and
$k_{\text{up}} = 1.5$.  These values will be used in all further simulations of
system (I), unless stated otherwise. The dynamics of the system towards the
steady state for diffusive exchange $k_d = 0.015$ is shown in
Fig.~\ref{fig:system_one} (b) and leads to a steady state with a dominance of
the non-toxic species, see Tab.~\ref{tab:steady_state}. As explained
previously the community structure in the system (I) directly reflects the low
amount of nutrients of this scenario. Please note that for $k_d = 0.015$ the
presence of zooplankton allows for the coexistence of the two phytoplankton
species. On the other hand, for a high nutrient supply $k = k_{\text{up}} = 1.5$
the dynamics leads to the dominance of the toxic species and even the extinction
of its competitor, see Fig.~\ref{fig:system_one} (c).

The parameters chosen for the further analysis for system (II) are:
$k_d = 0.0045$ and $k_{\text{up}} = 2.25$. Again we find, that with a large
input of nutrients the toxic species dominates, see
Tab.~\ref{tab:steady_state}. However, the main mechanism how this dominance is
achieved differs from previous case as explained previously.  Please note that
here we observe a significant amount of the total biomass concentrated at the
higher trophic level, especially in the low nutrient limit. Another distinction
of this set-up is the presence of both species of phytoplankton in the steady
state for low as well as high nutrient influx Fig.~\ref{fig:system_one} (e, f).
\begin{table}
  \begin{center}
 \begin{tabular}{|p{5em} c c|} 
   \hline
            &  Sys.(I) &  Sys.(II) \\[0.5ex]
   \hline\hline
   $k_d$ & $ 0.015$& $0.0045$ \\ [1.5ex] 
   $N$ & $ 0.02$& $0.034$ \\ [1.5ex] 
   $P_N$ & $0.04$ &  $0.007$  \\[1.5ex] 
   $P_T$ & $ 0.024$& $0.006$ \\ [1.5ex] 
   $Z$ & $0.028$ &  $0.016$  \\[1.5ex] 
 \hline
 \end{tabular}\caption{The steady state values in gC m$^{-3}$ for the systems subjected to $k_d$.}\label{tab:steady_state}
\end{center}
\end{table}

\subsection{Coupled Model}
The full biological-hydrodynamic model consists of the following
reaction-advection-diffusion system of equations:

\begin{equation}\label{eq:full_model}
 \frac{\partial C}{\partial t} + \vec{u} \nabla C = F_c + D \nabla^2 C, \qquad \text{with} \qquad C \in [N, P_N, P_T, Z],
\end{equation}

where $C(x, y,t)$ represents the concentration of nutrients or plankton species
in space and time, and $F_c$ are functions representing the biological
interactions among these species, which are given by the population dynamical
model Eq.(\ref{eq:pop_model}).  We consider a horizontal turbulent diffusion
constant $D = 10$ m$^2s^{-1} $, that describes the advection by smaller scales
in the flow field.  Please note that Eq.(\ref{eq:pop_model}) describes the
dynamics as vertically averaged model only in the mixed layer while vertical
transport is encapsulated in the biological model considering only the vertical
exchange of nutrients and the sinking of phytoplankton
(cf. subsection~\ref{sec:bio_model}).  The influx concentrations at the left
boundary are setup as $20\%$ of the steady state values of
Table~\ref{tab:steady_state}. For numerical details please check Supplemental
material. The code for the simulation reported here can be found in the Github
repository: https://github.com/kseniaguseva/Upwelling.

\section{Results}

\subsection{Hydrodynamic time scales}\label{sec:exit}
According to our aim to understand the conditions for HAB formation in the
presence of an intermittent upwelling, it is important to analyze the interplay
between hydrodynamic and biological time scales. We start the study of this
nontrivial coupling by analysing the hydrodynamics that underlies all the
biological processes in our system. To that extend we follow the motion of
non-reacting fluid parcels passively transported by the flow field (tracers).
Since we are interested in the impact of upwelling we compute the residence
times of tracers starting in the upwelling region. Furthermore, we want to
understand how the residence times of tracers depend on the initial time instant
of their release. We measure it by releasing the tracers at a location
($x_i, y_i$) and recording the time $\tau_i$ when they reach the right boundary
at $x = 8$.

\begin{figure}[h!]
\centering
\includegraphics[scale=1.]{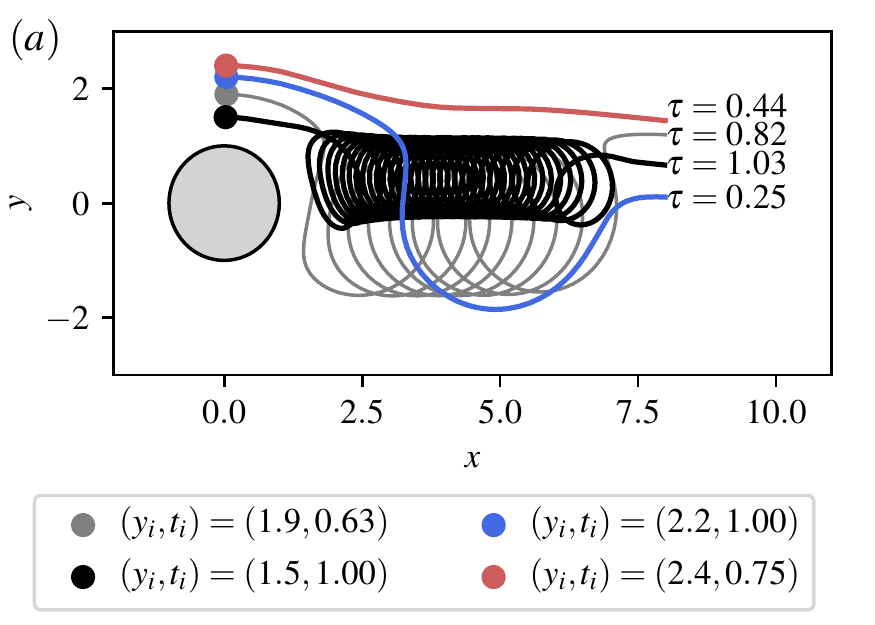}\\
\includegraphics[scale=0.8]{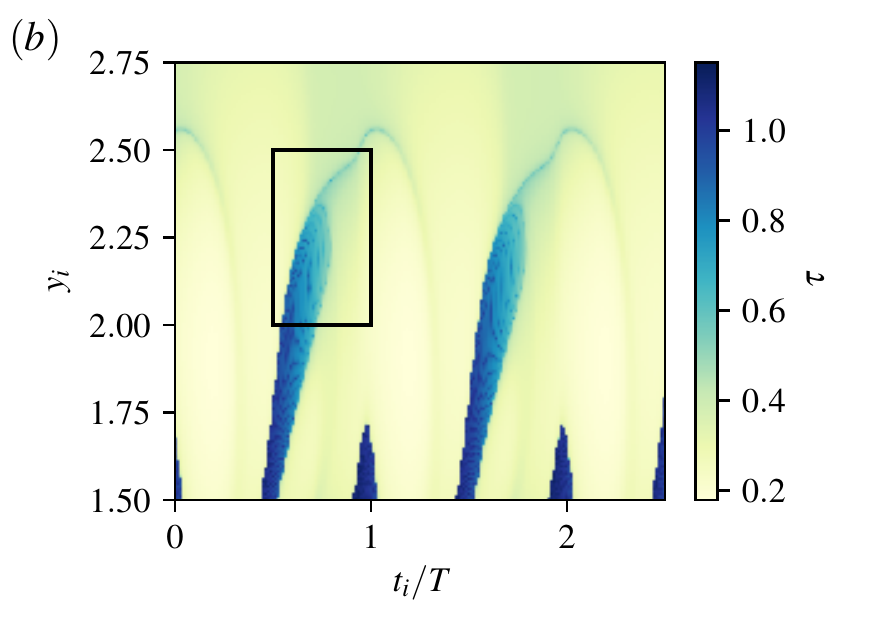}
\includegraphics[scale=0.8]{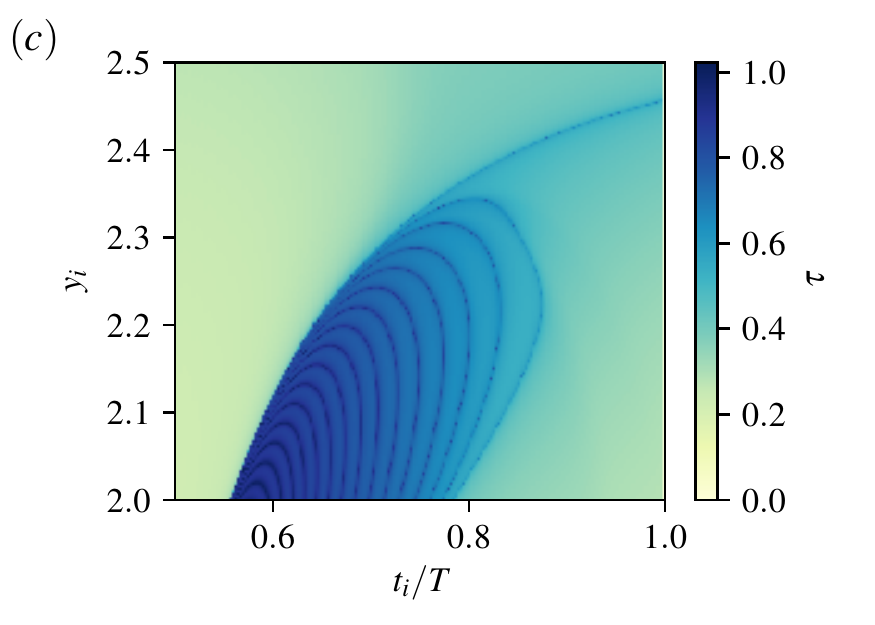}
\caption{(a) Typical trajectories of tracers that start in the
  upwelling region. (b) Residence times of tracers as a function of initial time
  $t_i$ and the vertical coordinate $y_i$. (c) amplification of the rectangular
  area of (b).}\label{fig:residence}
\end{figure}

We start by characterizing the possible trajectories that a tracer element can
take depending on its release time $t_i$ and its release coordinates
($x_i, y_i$), see typical trajectories and the respective $\tau_i$ in
Fig.~\ref{fig:residence} (a). The main difference in $\tau_i$ arises from
whether the tracer is captured by a vortex in the wake or not: The ones captured
into a circular trajectory around the vortex core (black and gray trajectories
in Fig.~\ref{fig:residence} (a)), spend at least two times longer in the
observation area than the ones that are transported more or less straight by the
main flow (red and blue trajectories of Fig.~\ref{fig:residence}
(a)). Fig.~\ref{fig:residence} (b) summarizes our results on residence times of
trajectories starting at ($x_i = 0, y_i, t_i$). The periodicity of the flow can
be seen by the repeating patterns shown in Fig.~\ref{fig:residence} (b). The two
finger-like structures in the residence times (blue points) correspond to
tracers that are captured by the vortices. Another important characteristics of
these patterns is their fractality (see Fig.~\ref{fig:residence}(c)), which
directly reflects the influence of the stable and unstable manifolds of the
chaotic saddle present in the system ~\cite{jung_application_1993,
  bastine_inhomogeneous_2010}.  Note that although we have chosen to fix $x_i$
at $x_i = 0$, the results for other release positions within the upwelling
region are very similar.

In summary, when we identify the tracers with nutrients released during
upwelling, then the residence time of nutrients in the observation region
changes with the position of the upwelling region, the location of release
within that region, and the time instant when the nutrients are released. In
other words, for how long nutrients, that have been released into the mixed
layer during upwelling, are available for consumption by phytoplankton in the
observation area depends crucially on the structure of the hydrodynamic flow at
the time instant of upwelling. Next we will investigate the consequences that
this effect has on the plankton dynamics.

\subsection{Reaction to an upwelling pulse: spatio-temporal patterns}\label{sec:spacetime}
Before we start with the results of this section we shortly discuss the
characteristics of the system in the absence of upwelling. In the absence of
upwelling the spatial distribution of the biological species follow the uneven
nutrient distribution in space. What is observed is an accumulation of nutrients
in certain areas, the observed accumulation appears due to the small advective
velocities around the island coupled with a constant nutrient influx from below
the thermocline ($k_d$). Subsequently, this high nutrient concentration is
captured by the vortices behind the island. This results into a bloom of
non-toxic species in these regions in System (I), and a non-toxic bloom followed
by the growth of zooplankton in System (II), see Supplemental material for
details. We define the values of the spatial average over the observation region
as $\left<C\right>$, where C stands for $N$, $P_N$, $P_T$ or $Z$. In addition,
we will also use a distinct notation for the time average of $\left<C\right>$
for the case without upwelling, defining it as
$\left<C\right>^* = \frac{1}{nT}\int_{0}^{nT} \left<C\right>dt$, the values for
the two systems are shown in Table~\ref{tab:av}.

\begin{table}
  \begin{center}
 \begin{tabular}{|p{5em} c c|} 
   \hline
            &  Sys.(I) &  Sys.(II) \\[0.5ex]
   \hline\hline
   $\left<N\right>^{*}$ & $ 0.025$ & $0.0286$ \\ [1.5ex] 
   $\left<P_N\right>^{*}$ & $0.042$ &  $0.0084$  \\[1.5ex] 
   $\left<P_T\right>^{*}$ & $0.019$& $0.0010$ \\ [1.5ex] 
   $\left<Z\right>^{*}$ & $0.005$ &  $0.0054$  \\[1.5ex] 
 \hline
 \end{tabular}\caption{Spatial averages in gC m$^{-3}$ for the system without upwelling.}\label{tab:av}
\end{center}
\end{table}

After having analysed the coupled hydrodynamic-biological model, let us
characterize the HAB formed in the two systems in the presence of a simple
upwelling pulse. We start by comparing the spatio-temporal patterns for the two
biological systems for a case where the upwelling event triggers a HAB.  We
introduce a single upwelling event at $2.5$ T, at this instant the value $k_d$
at the upwelling area is exchanged to $k_{\text{up}}$. It is kept at this
constant value for some time interval $\delta = 0.5$, and then it is exchanged
back to $k_d$.

\begin{figure}[h!]
\includegraphics[scale=.8]{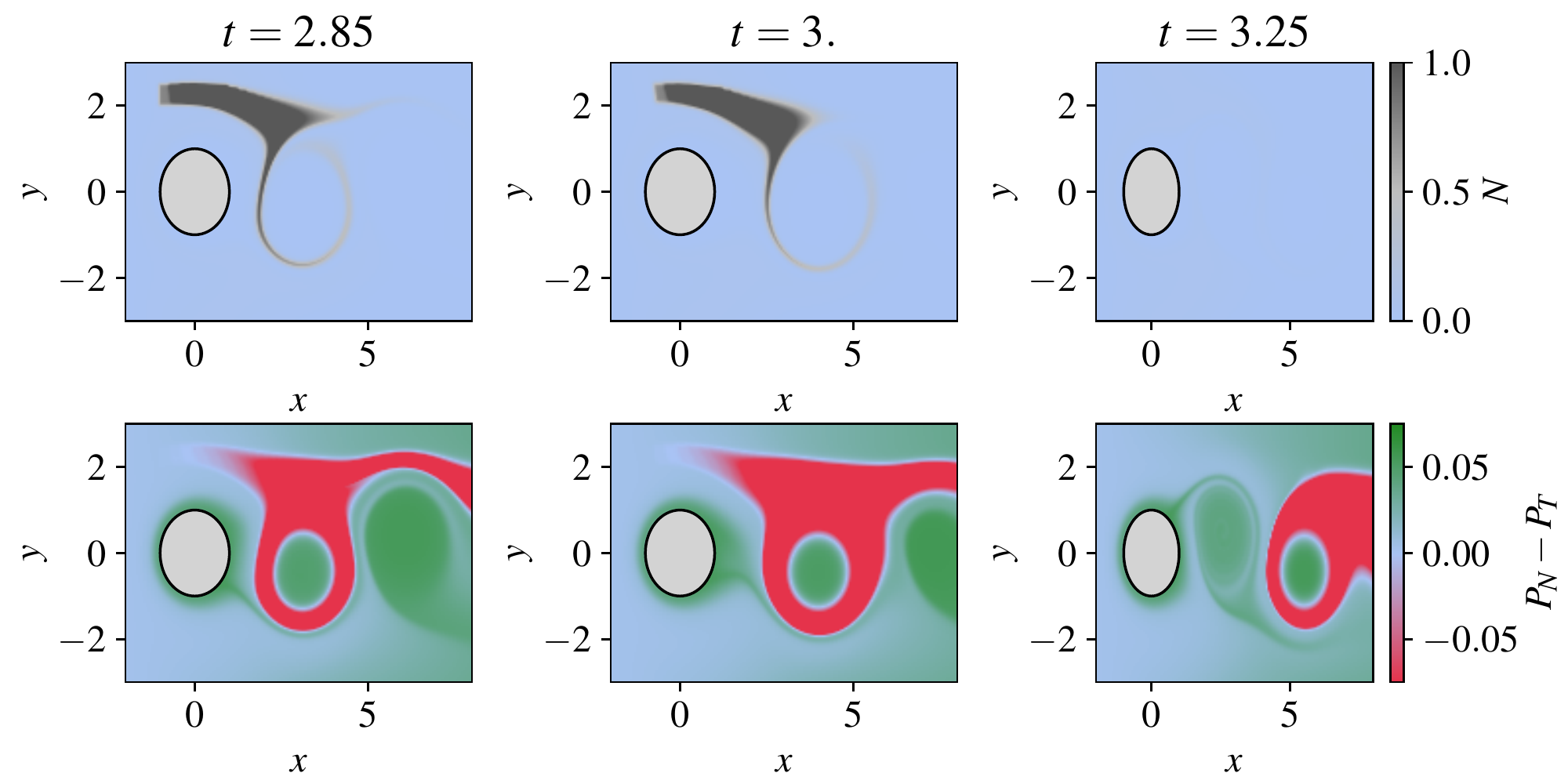}
\caption{Spatial distribution of the dominance patterns which follow an
  upwelling event at $2.5$ T.}\label{fig:puls1_space}
\end{figure}

In system (I) the spatio-temporal distribution is simple: the non-toxic species
grows in the vortex cores undisturbed by the upwelling event, while the toxic
species grows mainly by feeding on nutrients released by the upwelling.
Fig.~\ref{fig:puls1_space} illustrates these dominance patterns for three
instances of time that follow an upwelling event. Here we have specifically
chosen an upwelling event that triggers a strong dominance change. Please note
the fast reaction of the toxic species to the nutrient influx.

\begin{figure}[h!]
\includegraphics[scale=.8]{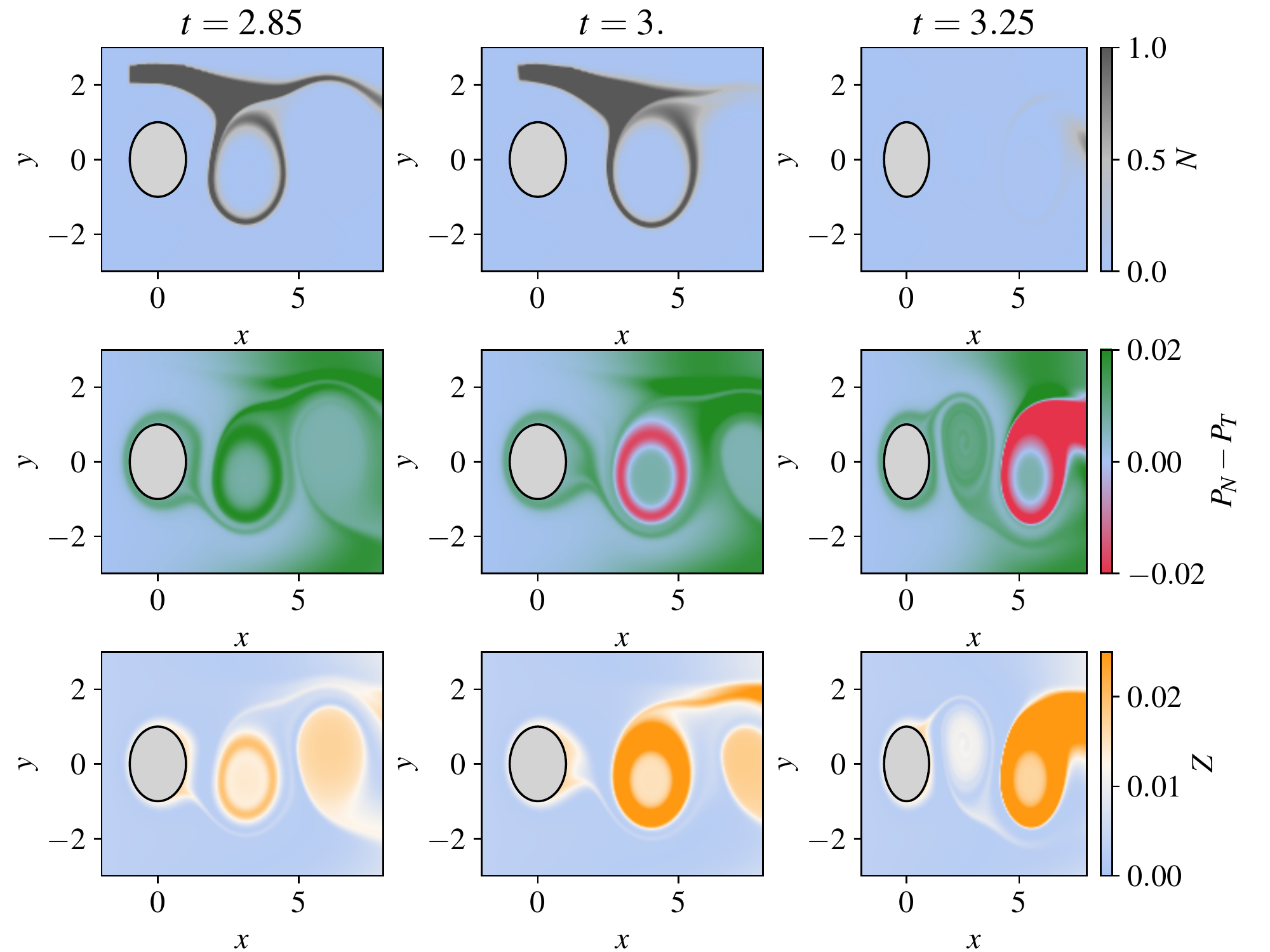}
\caption{Spatial distribution of the dominance patterns which follow an
  upwelling event at $2.5$ T. }\label{fig:puls2_space}
\end{figure}

By contrast, in system (II) both phytoplankton species readily grow in response
to the nutrients. However there is a stronger response of the non-toxic species
due to its lower half saturation constant, which allows it to reach high
concentration and initiate the growth of zooplankton. However, the zooplankton
development is a slow process and it only reaches significant concentrations
when the non-toxic bloom is captured by a vortex. It is in this region where the
toxic species, with extra nutrients and the presence of zooplankton, can
successfully compete with non-toxic species. In fact, the high grazing pressure
of zooplankton on the non-toxic species allows for the very localized dominance
of the toxic specie, see Fig.~\ref{fig:puls2_space}. Note that when the bloom of
the toxic species forms, the nutrients brought by the upwelling were already
partially consumed.

We would like to emphasize that the spatio-temporal dominance patterns that
appear in this system in the presence of upwelling in system (I) and (II)
strongly differ. This difference can be explained by the fact, that the two
spatio-temporal patterns result from distinct biological mechanisms. The
behaviour in system (I) is solely determined by the bottom up control relying
only on the supply of nutrients leading to a strong advantage of the toxic
species in areas of high nutrient concentration. By contrast, in system (II) the
top down control by the zooplankton is the dominant biological process shaping
the spatio-temporal pattern. The toxic species can only dominate in areas, where
its competitor is kept at low concentration due to the high grazing
pressure. Furthermore, the two set-ups are characterised by different response
times of the toxic species to the inflow of nutrients in the two systems.

\subsection{The impact of initial time of the upwelling event}\label{sec:init}

\begin{figure}[h!]
  \centering
  \includegraphics[scale=.9]{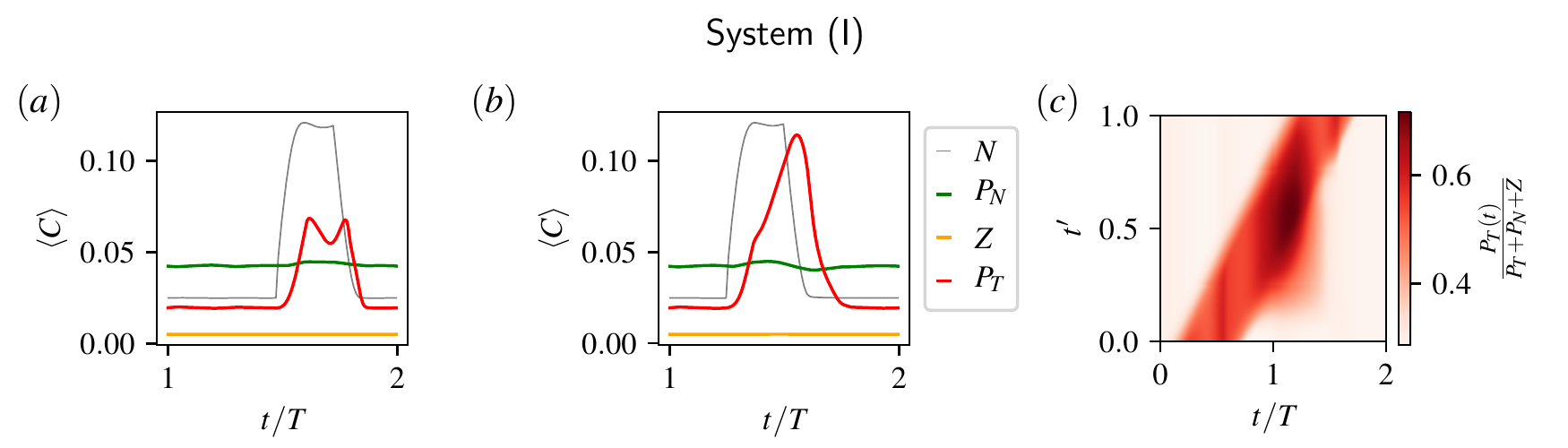}
  \includegraphics[scale=.9]{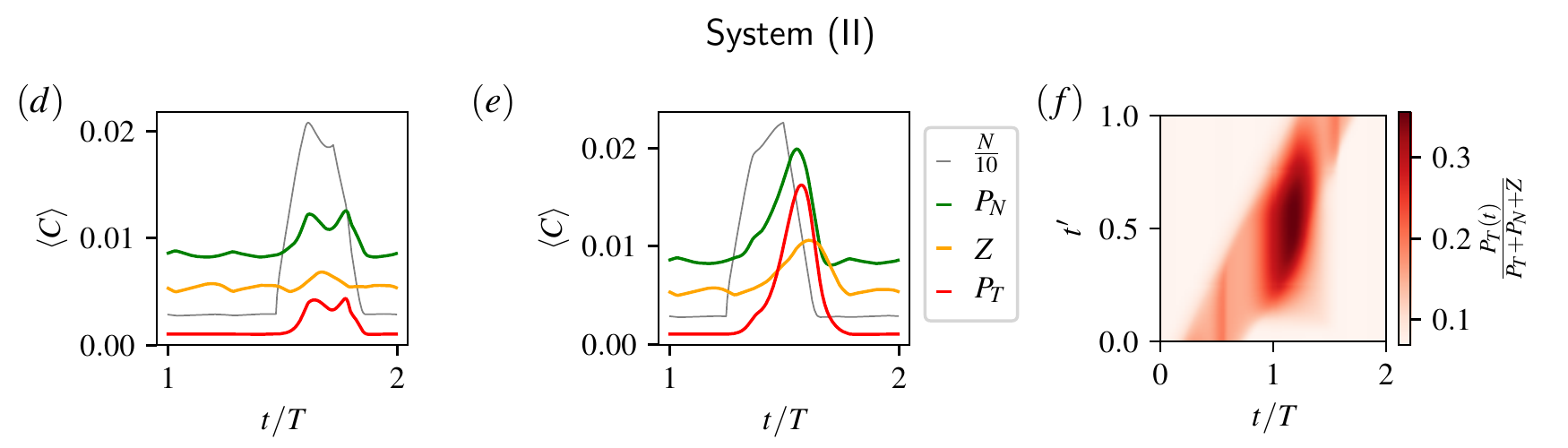}
  \caption{(a,b, d, e) Two possible responses of the population dynamics to the
    influx of nutrients through upwelling events ($\delta = 0.5$) initiated at:
    (a, d) $t' = 0.95$ and (b, e) $t' = 0.5$. (c, f) The time evolution of the
    relative biomass of the toxic species following upwelling events. The top
    panel (a, b, c) represent the results for System (I) and the bottom panel
    (d, e, f) for System (II).}\label{fig:zoom_puls}
\end{figure}

In Sec.~\ref{sec:exit} we have illustrated that fluid parcels released at
different times, $t_i$, from the upwelling region can take very distinct paths
trough the observation area. Some of these paths transport the fluid parcels
directly away, describing a quick escape from the observation area, while others
consists of spiral trajectories around vortex cores. These latter trajectories,
in turn, are characterized by long residence times. In this section we will
connect the advection with the plankton dynamics. Our objective is to answer how
these different time scales affect the formation of HABs. Thus, we initialize
upwelling pulses starting from different initial times $t'$, with a predefined
duration $\delta$ and strength $k_{\text{up}}$.

Now we analyse how these upwelling events impact the time series of the spatial
averages of the plankton species of our biological model.  While in the time
series of system (I) only the toxic species exhibits a strong response to the
upwelling events, in system (II) we observe, on the contrary, spikes in the
growth of both phytoplankton populations and even in the abundance of
zooplankton (Fig.~\ref{fig:zoom_puls} (a,d)). Despite these differences, we
observe that in both systems the dynamics of the response of the plankton model
to the upwelling event depends on its initial time $t'$: in both systems we can
have weak or strong responses, see Fig.~\ref{fig:zoom_puls} (a, d) and (b,e)
respectively. This result is summarised in Fig.~\ref{fig:zoom_puls} (c,f) where
the biomass of the toxic species is compared to the total biomass for an event
with duration $\delta = 0.5$. While for system (I) which is solely nutrient
controlled we observe a dominance change for the average concentrations, this
behaviour is absent for system (II), which has a strong top down control
element. In system (II) we observe only local dominance change which never
reaches a dominance of the toxic species in the spatial average. Please note the
similarity of the diagrams of the two biological systems. The similarity of the
response patterns for both systems (I) and (II) with respect to the timing of
the response is entirely determined by the hydrodynamics.

\subsection{Impact of the duration and strength of the upwelling
  event}\label{sec:duration}

So far we have seen that the initial time of an upwelling event plays a crucial
role for the mechanism of formation of HABs.  In this section we extend our
analysis to investigating the effect of the duration and the upwelling strength
of randomly initialized upwelling events.

\begin{figure}[h!]
  \centering
  \includegraphics[scale=1.]{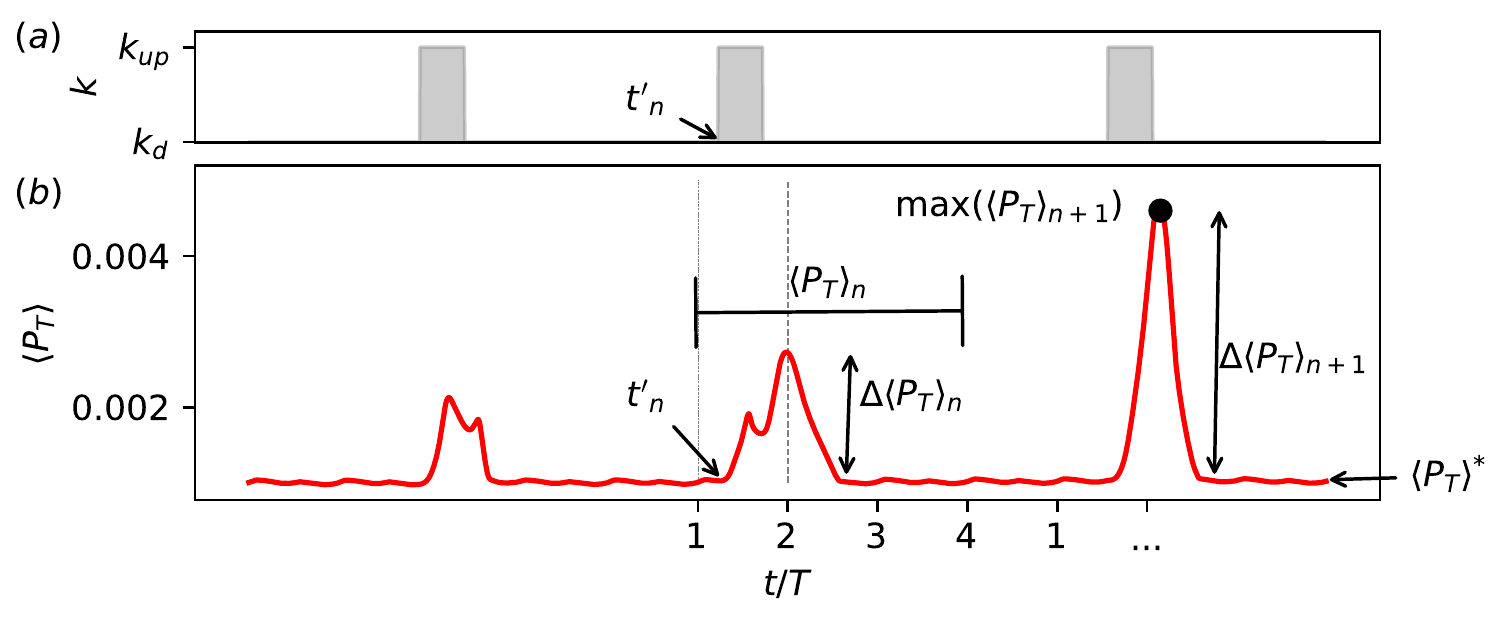}  
  \caption{ (a) Time series of the strength of the
    thermocline exchange rate at the upwelling region; (b) Example of
    a time series of the spatial average of concentration of the toxic species,
    with blooming events triggered by upwelling.}\label{fig:scheme_pulse}
\end{figure}

We compose the sequence of upwelling pulses in the following way: The upwelling
events are initiated at particular time instants given by $4Tn +t'_n$, where
$n \in$ $\Bbb N^{*}$ and $t'_n$ is chosen randomly for every $n$ from the
interval $[0, T]$, see Fig.~\ref{fig:scheme_pulse} (a). We establish that each
sequence is characterized by upwelling events of duration $\delta$ and strength
$k_{\text{up}}$. To quantify the effect of the upwelling on the growth of the
phytoplankton species, the time series is divided into $n$ intervals: each one
of them containing four periods and a single upwelling event.  The time series
of the average concentration $\left<C\right>$ for each one of these intervals is
denoted $\left<C\right>_n$. Thus, the effect of each upwelling event on the
population dynamics is reflected in the maximum,
max$\left(\left<C\right>_n\right)$. Furthermore it is useful to systematically
compare max$\left(\left<C\right>_n\right)$ to the average concentration in the
absence of upwelling, $\left<C\right>^*$, we represent this deviation by
$\Delta \left< C \right>_n= $ max$(\left<C\right>_n) - \left<C\right>^*$ (see
Fig.~\ref{fig:scheme_pulse}).

\begin{figure}[h!]
  \centering
  \includegraphics[scale=.9]{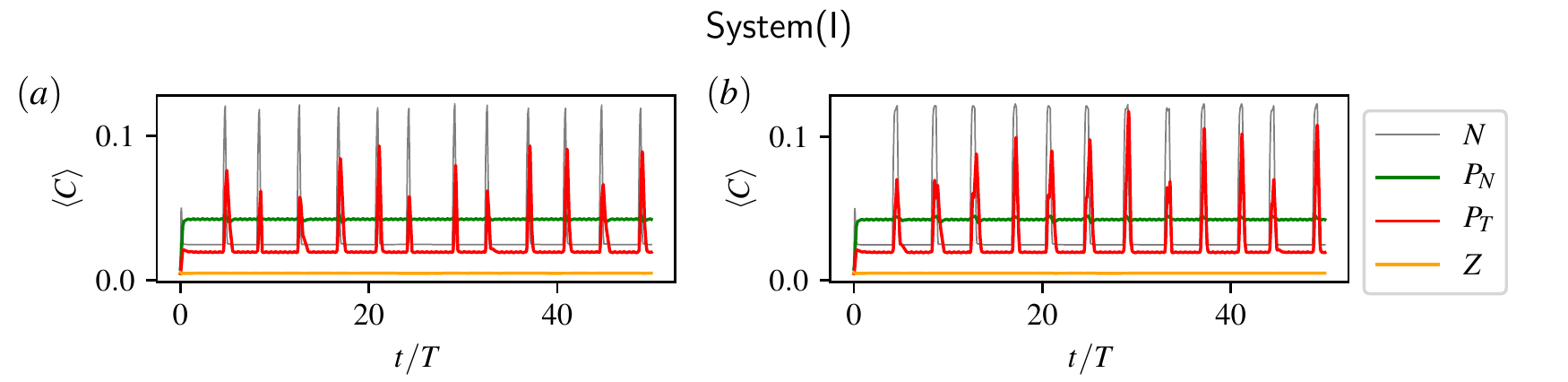}
  \includegraphics[scale=.9]{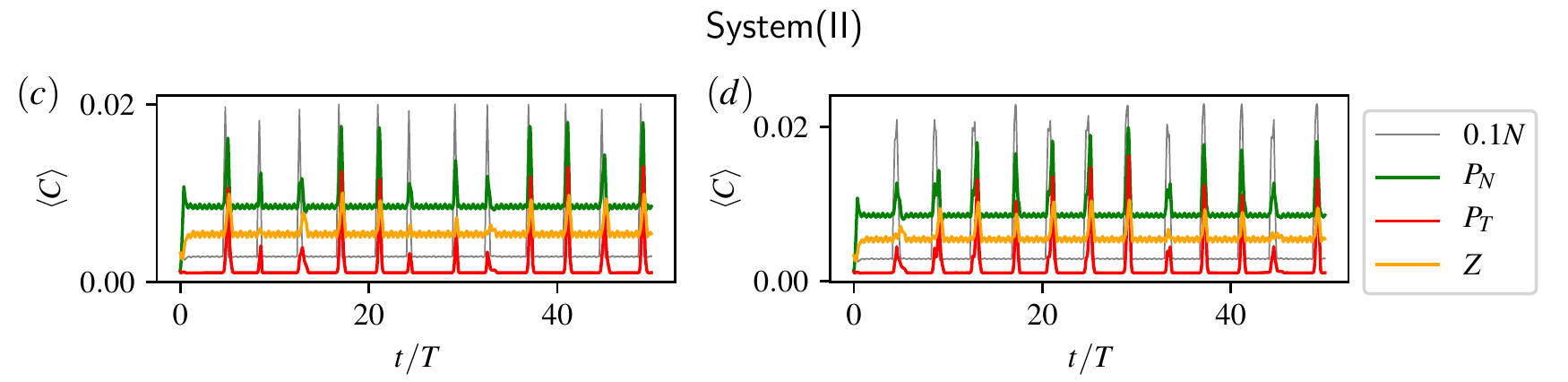}
  \caption{Time series of the spatial averages of the concentrations of our
    biological model $\left<C\right>$ in (a, b) system(I) and (c, d) system
    (II).  The systems are subjected to randomly initiated pulses of upwelling
    characterised by the duration: (a, c) $\delta = 0.25$ T; (b, d)
    $\delta = 0.5$ T.}\label{fig:puls1_tseries}
\end{figure}

\begin{figure}[h!]
  \centering
  \includegraphics[scale=1.2]{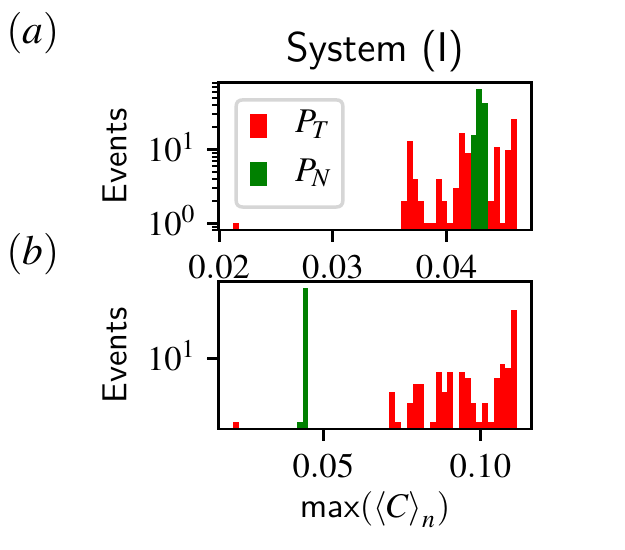}
  \includegraphics[scale=1.2]{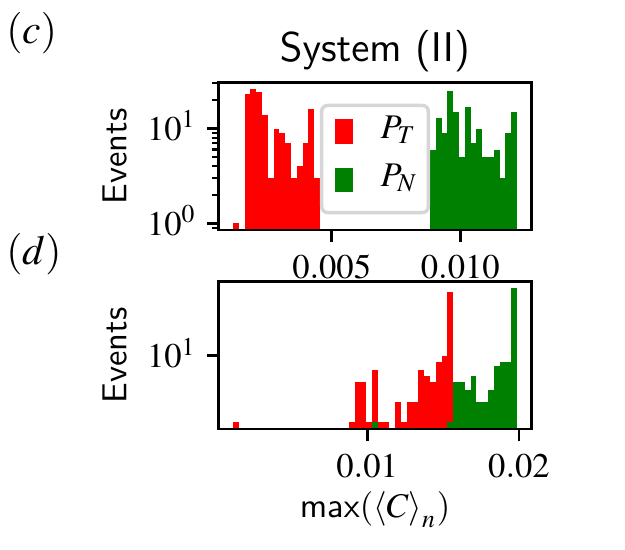}
  \caption{ The distribution of the observed responses of the toxic and
    non-toxic species to the upwelling pulses, $250$ upwelling events
    distributed over $1000$ T for (a, b) System (I) and (c, d) System (II)
    using: (a, c) $k_{\text{up}} = 1.0$ and $\delta = 0.08$; (b, d)
    $k_{\text{up}} = 1.0$ and $\delta = 0.95$.}\label{fig:puls1_stat}
\end{figure}

We start our analysis by fixing $k_{\text{up}}$. In the resulting time series of
system (I), see Fig.~\ref{fig:puls1_tseries} (a, b), the toxic species is the
only species that shows a response to upwelling in its average values. On the
contrary, in system (II) all the species show a bloom-like behaviour,
Fig.~\ref{fig:puls1_tseries} (c, d). It is clear that the average values shown
for both of these systems, fail to completely describe the complexity of the
spatio-temporal dynamics. Nevertheless, part of this complexity is revealed by
the variability of different dynamic responses of the biological community to
seemingly identical upwelling events, see
Fig.~\ref{fig:puls1_tseries}. Comparing the different responses for the same
system with the same duration, we notice that it depends crucially on the timing
of the upwelling event, how strong the response is going to be. This revels
clearly the importance of the structure of the flow field at the time instant of
the upwelling. Additionally, our results reveal that this variability depends on
the duration $\delta$, and this relation manifests itself in a similar way for
both systems (Fig.~\ref{fig:puls1_tseries}). Our results reveal that longer
upwelling events are associated with a vigorous growth of the toxic species. For
this case the probability of HABs is large and we observe similar peaks in the
concentration of the toxic species (large values of
$\Delta \left< P_T \right>_n$). On the other hand, shorter $\delta$ values
reveal a larger variety of possible outcomes. These results are summarized in
the histograms of Fig.~\ref{fig:puls1_stat} (note the difference of the axes
between the upper and the lower panels). The observed behaviour can be explained
by taking into account that HAB formation depends on the temporal overlap
between the upwelling event and the vortex formation in the wake of the
island. Naturally for larger $\delta$ the probability of this overlap is higher
and more nutrients are captured to incubate the growth of the toxic species. The
strength of upwelling events has a complementary influence. For small values of
$k_{\text{up}}$ the system needs longer upwelling events to release enough
nutrients for toxic species bloom, see the Supplemental material for details.

At the end of this section we want to stress that from an analysis of the time
series only, it is especially difficult to establish a causal relation between
the upwelling event and the rate of increase of the toxic species. Although, an
increase in the population of the toxic species always follows the upwelling in
our model system, the level of increase in the population varies strongly, see
Fig.~\ref{fig:puls1_stat} (d). This variety of the possible outcomes, however,
can be easily explained by coupling the biological model with hydrodynamic
mesoscale motion. Therefore by taking into account the interplay between the
initial time of the upwelling event and the formation of vortices in the wake,
it is possible to predict if the event will result in a HAB formation.

\subsection{Intermittent upwelling events}\label{sec:inter}

In the previous sections we have seen that an upwelling pulse, even of the
simplest possible profile, can result in a variety of possible outcomes for the
plankton growth. The intricate interplay between plankton dynamics and the
formation of vortices, or more general mesoscale hydrodynamic structures,
results in time series showing responses of {\it different} strengths for {\it
  identical} upwelling events. Here in this section we analyse the response of
our model to upwelling events that follow a time series that displays more
complex patterns. The idea here is to mimic a more realistic situation, since
upwelling is a wind driven phenomenon and hence, has an intermittent character.
To generate this new time series of upwelling events we use a dynamical system
which displays a special type of intermittent behaviour, known as ``on-off''
intermittency~\cite{platt_-off_1993}. Two modes appear in this system: the
``off'' mode (situation without upwelling) where a very small value of an
observable of the system sets up for long intervals of time; these intervals are
interrupted by seemingly random bursts, characteristic to the ``on'' mode
(upwelling events). Therefore the thermocline exchange rate at the upwelling
region in the ``off'' mode is $k_{d}$ and in an ``on'' mode $k_{\text{up}}$,
which here assumes a set of random values obtained from the dynamical system
described in the Supplemental material.

\begin{figure}[h!]
  \includegraphics[scale=.8]{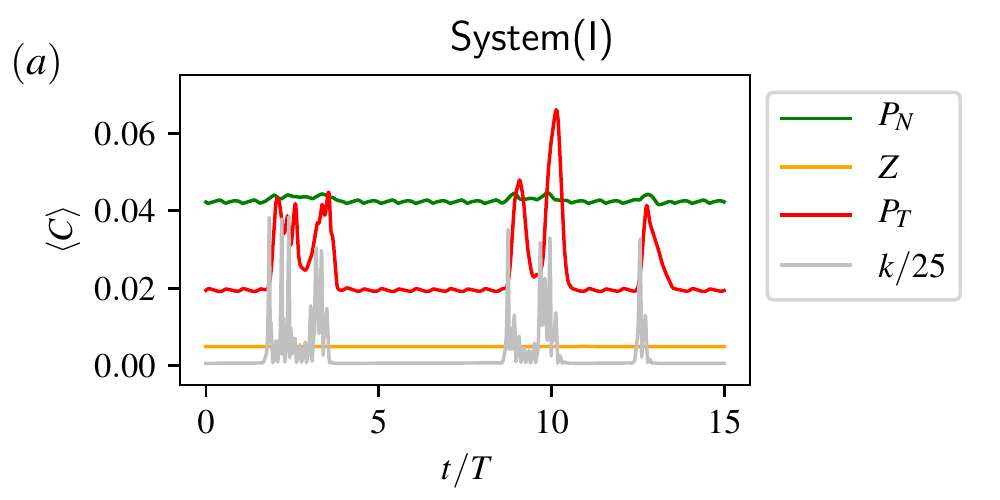}
  \includegraphics[scale=.8]{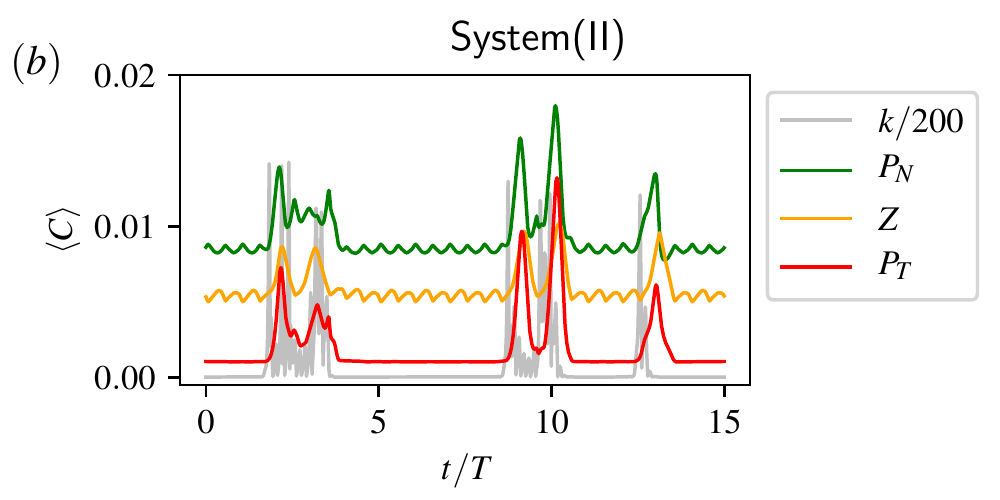}
  \caption{Population dynamics in response to an intermittent pulses of
    upwelling for (a) System (I) and (b) System (II).}\label{fig:interm_1}
\end{figure}

Fig.~\ref{fig:interm_1} shows the response of our two biological systems to an
identical sequence of intermittent upwelling pulses shown in light gray
Fig.~\ref{fig:interm_1}. Note that in this system there can be several short
pulses of different strengths within a single period, furthermore the events are
not isolated but come in small groups. Each group of pulses triggers a different
outcome for the toxic population. Furthermore, as can be easily spotted in
Fig. ~\ref{fig:interm_1}, there is a very strong variability between possible
responses. Note, for instance the weak blooming behaviour of the toxic species
around $t = 2.5$ in contrast to the strong response at $t \sim 10$.  For system
(I) we find at $t \sim 2.5$ a rather long bloom with moderate amplitude, while
at $t \sim 10$ the bloom exhibits a much higher amplitude. For system (II) we
find a similar response, but now not only for the toxic but also for the
non-toxic one and the zooplankton reflecting the importance of the grazing
pressure in that case.

\section{Conclusion}\label{sec:conclusion}

In this work, we have analysed how the competition, between two species for a
shared limiting resource, can be affected by intermittent upwelling events
providing an additional input of this resource. We have used a theoretical
approach which couples the hydrodynamic flow field with a biological model by
means of reaction-advection-diffusion equations. Notably, we have tracked the
necessary environmental conditions that trigger a HAB. We were particularly
interested on how the interplay of the hydrodynamic timescales as well as the
mesoscale hydrodynamics structures, like vortices in the flow, coupled to
intermittent upwelling pulses influence the spatio-temporal distribution of
dominance patterns of different functional groups of phytoplankton. First we
have characterized the HAB formation in two biological scenarios: the first
scenario where the abundance of nutrients is the only factor responsible for the
emergence of dominance patterns in the system; and the second one, where the
dominance patterns arise from combination of competition for nutrients and
grazing pressure from a higher trophic level.  Both scenarios are characterized
by distinct spatio-temporal inhomogeneous distributions of the phytoplankton
groups, which appear as a result of an upwelling event. In the first scenario
the toxic species develops along the whole nutrient plume, while in the second
system a bloom is formed in a very localized region namely on a narrow ring
around one of the vortices in the wake. The time of the bloom development also
differs in these systems: in the first one the response of the toxic species is
almost immediate, while in the second one the dominance change occurs while the
vortex is advected away from the island. Despite the observed differences in
theses two systems we demonstrate that, in both of them, the decisive factor
triggering a bloom or not is the coupling of the upwelling event with the
formation of mesoscale vortices. In this scenario the HAB formation results from
the interplay of three timescales: (1) of the vortex formation at the island's
wake, (2) of the upwelling event and (3) of the biological growth. Our analysis
shows that identical upwelling events that start in different instances of time
may result in a variety of outcomes for the biological community depending on
the properties of the flow at the moment of upwelling. The observed response
depends on the time interval that nutrients released by upwelling spend in the
observation region and consequently the quantity of nutrients captured by the
vortices. Therefore the variability of these possible outcomes depends also on
the duration and the strength of the upwelling events.

In summary we have observed that the HAB formation, independently of the
biological set-up, is tightly associated with the transport dynamics of the flow
field. From our analysis we conclude that without taking advection into account
it appears to be not possible to establish the relationship between upwelling
events and triggering a HAB. For this reason one cannot expect to find a
functional dependence between upwelling events and plankton blooms in general,
when only nutrients and plankton abundances are measured and no information
about the flow field is available. Such measurements lacking the properties of
the flow field will always be difficult to interpret and allow only conclusions
when the flow field is simple and does not contain mesoscale hydrodynamics
structures.

\section*{Acknowledgments}
We are grateful to Rahel Vortmeyer-Kley for illuminating discussions.
This work was supported by the Volkswagen Foundation (Grant No. 88459).

\bibliographystyle{vancouver}
\bibliography{Upwelling}

\begin{thebibliography}{10}

\bibitem{pettersson_monitoring_2013}
Pettersson LH, Pozdnyakov D.
\newblock Monitoring of {Harmful} {Algal} {Blooms}.
\newblock Geophysical {Sciences}. Berlin Heidelberg: Springer-Verlag; 2013.

\bibitem{mann_dynamics_2005}
Mann K, Lazier J.
\newblock Dynamics of {Marine} {Ecosystems}: {Biological}-{Physical}
  {Interactions} in the {Oceans}.
\newblock Wiley; 2005.

\bibitem{belgrano_north_1999}
Belgrano A, Lindahl O, Hernroth B.
\newblock North {Atlantic} {Oscillation} primary productivity and toxic
  phytoplankton in the {Gullmar} {Fjord}, {Sweden} (1985–1996).
\newblock Proceedings of the Royal Society of London B: Biological Sciences.
  1999 Mar;266(1418):425--430.

\bibitem{sellner_harmful_2003}
Sellner KG, Doucette GJ, Kirkpatrick GJ.
\newblock Harmful algal blooms: causes, impacts and detection.
\newblock Journal of Industrial Microbiology and Biotechnology. 2003
  Jul;30(7):383--406.

\bibitem{kahru_ocean_2008}
Kahru M, Mitchell BG.
\newblock Ocean {Color} {Reveals} {Increased} {Blooms} in {Various} {Parts} of
  the {World}.
\newblock Eos, Transactions American Geophysical Union. 2008;89(18):170--170.

\bibitem{martin_phytoplankton_2003}
Martin AP.
\newblock Phytoplankton patchiness: the role of lateral stirring and mixing.
\newblock Progress in Oceanography. 2003 May;57(2):125--174.

\bibitem{gower_phytoplankton_1980}
Gower JFR, Denman KL, Holyer RJ.
\newblock Phytoplankton patchiness indicates the fluctuation spectrum of
  mesoscale oceanic structure.
\newblock Nature. 1980 Nov;288(5787):157--159.

\bibitem{mcgillicuddy_mechanisms_2016}
McGillicuddy DJ.
\newblock Mechanisms of {Physical}-{Biological}-{Biogeochemical} {Interaction}
  at the {Oceanic} {Mesoscale}.
\newblock Annual Review of Marine Science. 2016;8(1):125--159.

\bibitem{levy_modulation_2008}
Levy M.
\newblock The {Modulation} of {Biological} {Production} by {Oceanic}
  {Mesoscale} {Turbulence}.
\newblock In: Weiss JB, Provenzale A, editors. Transport and {Mixing} in
  {Geophysical} {Flows}: {Creators} of {Modern} {Physics}. Lecture {Notes} in
  {Physics}. Berlin, Heidelberg: Springer Berlin Heidelberg; 2008. p. 219--261.

\bibitem{lehahn_satellite-based_2018}
Lehahn Y, d'Ovidio F, Koren I.
\newblock A {Satellite}-{Based} {Lagrangian} {View} on {Phytoplankton}
  {Dynamics}.
\newblock Annual Review of Marine Science. 2018;10(1):99--119.

\bibitem{mackas_spectral_1979}
Mackas DL, Boyd CM.
\newblock Spectral {Analysis} of {Zooplankton} {Spatial} {Heterogeneity}.
\newblock Science. 1979 Apr;204(4388):62--64.

\bibitem{martin_plankton_2002}
Martin AP, Srokosz MA.
\newblock Plankton distribution spectra: inter-size class variability and the
  relative slopes for phytoplankton and zooplankton.
\newblock Geophysical Research Letters. 2002 Dec;29(24):66--1--66--4.

\bibitem{weber_variance_1986}
Weber LH, El-Sayed SZ, Hampton I.
\newblock The variance spectra of phytoplankton, krill and water temperature in
  the {Antarctic} {Ocean} south of {Africa}.
\newblock Deep Sea Research Part A Oceanographic Research Papers. 1986
  Oct;33(10):1327--1343.

\bibitem{abraham_generation_1998}
Abraham ER.
\newblock The generation of plankton patchiness by turbulent stirring.
\newblock Nature. 1998 Feb;391(6667):577--580.

\bibitem{bracco_horizontal_2009}
Bracco A, Clayton S, Pasquero C.
\newblock Horizontal advection, diffusion, and plankton spectra at the sea
  surface.
\newblock Journal of Geophysical Research: Oceans. 2009 Feb;114(C2).

\bibitem{mckiver_influence_2009}
McKiver WJ, Neufeld Z.
\newblock Influence of turbulent advection on a phytoplankton ecosystem with
  nonuniform carrying capacity.
\newblock Physical Review E. 2009 Jun;79(6):061902.

\bibitem{mckiver_plankton_2009}
McKiver W, Neufeld Z, Scheuring I.
\newblock Plankton bloom controlled by horizontal stirring.
\newblock Nonlin Processes Geophys. 2009 Oct;16(5):623--630.

\bibitem{hernandez-garcia_sustained_2004}
Hernández-García E, López C.
\newblock Sustained plankton blooms under open chaotic flows.
\newblock Ecological Complexity. 2004 Sep;1(3):253--259.

\bibitem{dovidio_fluid_2010}
d’Ovidio F, Monte SD, Alvain S, Dandonneau Y, Lévy M.
\newblock Fluid dynamical niches of phytoplankton types.
\newblock Proceedings of the National Academy of Sciences. 2010
  Oct;107(43):18366--18370.

\bibitem{vortmeyer-kley_eddies:_2019}
Vortmeyer-Kley R, Lünsmann B, Berthold M, Gräwe U, Feudel U.
\newblock Eddies: {Fluid} {Dynamical} {Niches} or {Transporters}?–{A} {Case}
  {Study} in the {Western} {Baltic} {Sea}.
\newblock Frontiers in Marine Science. 2019;6.

\bibitem{levy_dynamical_2015}
Levy M, Jahn O, Dutkiewicz S, Follows MJ, d'Ovidio F.
\newblock The dynamical landscape of marine phytoplankton diversity.
\newblock Journal of the Royal Society Interface. 2015 Oct;12(111).

\bibitem{karolyi_chaotic_2000}
Karolyi G, Pentek A, Scheuring I, Tel T, Toroczkai Z.
\newblock Chaotic flow: {The} physics of species coexistence.
\newblock Proceedings of the National Academy of Sciences. 2000
  Dec;97(25):13661--13665.

\bibitem{scheuring_competing_2003}
Scheuring I, Karolyi G, Toroczkai Z, Tel T, Pentek A.
\newblock Competing populations in flows with chaotic mixing.
\newblock Theoretical Population Biology. 2003 Mar;63(2):77--90.

\bibitem{tel_chemical_2005}
Tel T, de~Moura A, Grebogi C, Karolyi G.
\newblock Chemical and biological activity in open flows: {A} dynamical system
  approach.
\newblock Physics Reports. 2005 Jul;413(2):91--196.

\bibitem{bracco_mesoscale_2000}
Bracco A, Provenzale A, Scheuring I.
\newblock Mesoscale vortices and the paradox of the plankton.
\newblock Proceedings of the Royal Society B: Biological Sciences. 2000
  Sep;267(1454):1795--1800.

\bibitem{rossi_comparative_2008}
Rossi V, López C, Sudre J, Hernández‐García E, Garçon V.
\newblock Comparative study of mixing and biological activity of the {Benguela}
  and {Canary} upwelling systems.
\newblock Geophysical Research Letters. 2008 Jun;35(11).

\bibitem{rossi_surface_2009}
Rossi V, López C, Hernández-García E, Sudre J, Garçon V, Morel Y.
\newblock Surface mixing and biological activity in the four {Eastern}
  {Boundary} {Upwelling} {Systems}.
\newblock Nonlin Processes Geophys. 2009 Aug;16(4):557--568.

\bibitem{gruber_eddy-induced_2011}
Gruber N, Lachkar Z, Frenzel H, Marchesiello P, Münnich M, McWilliams JC,
  et~al.
\newblock Eddy-induced reduction of biological production in eastern boundary
  upwelling systems.
\newblock Nature Geoscience. 2011 Nov;4(11):787--792.

\bibitem{sandulescu_kinematic_2006}
Sandulescu M, Hernández‐García E, López C, Feudel U.
\newblock Kinematic studies of transport across an island wake, with
  application to the {Canary} islands.
\newblock Tellus A. 2006;58(5):605--615.

\bibitem{sandulescu_plankton_2007}
Sandulescu M, López C, Hernández-García E, Feudel U.
\newblock Plankton blooms in vortices: the role of biological and hydrodynamic
  timescales.
\newblock Nonlin Processes Geophys. 2007 Aug;14(4):443--454.

\bibitem{sandulescu_biological_2008}
Sandulescu M, López C, Hernández-García E, Feudel U.
\newblock Biological activity in the wake of an island close to a coastal
  upwelling.
\newblock Ecological Complexity. 2008 Sep;5(3):228--237.

\bibitem{bastine_inhomogeneous_2010}
Bastine D, Feudel U.
\newblock Inhomogeneous dominance patterns of competing phytoplankton groups in
  the wake of an island.
\newblock Nonlin Processes Geophys. 2010 Dec;17(6):715--731.

\bibitem{trainer_domoic_2000}
Trainer VL, Adams NG, Bill BD, Stehr CM, Wekell JC, Moeller P, et~al.
\newblock Domoic acid production near {California} coastal upwelling zones,
  {June} 1998.
\newblock Limnology and Oceanography. 2000;45(8):1818--1833.

\bibitem{kudela_harmful_2005}
Kudela R, Pitcher G, Probyn T, Figueiras F, Moita T, Trainer V.
\newblock Harmful {Algal} {Blooms} in {Coastal} {Upwelling} {Systems}.
\newblock Oceanography. 2005 Jun;18(2):184--197.

\bibitem{omand_episodic_2012}
Omand MM, Feddersen F, Guza RT, Franks PJS.
\newblock Episodic vertical nutrient fluxes and nearshore phytoplankton blooms
  in {Southern} {California}.
\newblock Limnology and Oceanography. 2012;57(6):1673--1688.

\bibitem{bialonski_phytoplankton_2016}
Bialonski S, Caron DA, Schloen J, Feudel U, Kantz H, Moorthi SD.
\newblock Phytoplankton dynamics in the {Southern} {California} {Bight}
  indicate a complex mixture of transport and biology.
\newblock Journal of Plankton Research. 2016 Aug;38(4):1077--1091.

\bibitem{nezlin_phytoplankton_2012}
Nezlin NP, Sutula MA, Stumpf RP, Sengupta A.
\newblock Phytoplankton blooms detected by {SeaWiFS} along the central and
  southern {California} coast.
\newblock Journal of Geophysical Research: Oceans. 2012;117(C7).

\bibitem{martin_mechanisms_2001}
Martin AP, Richards KJ.
\newblock Mechanisms for vertical nutrient transport within a {North}
  {Atlantic} mesoscale eddy.
\newblock Deep Sea Research Part II: Topical Studies in Oceanography. 2001
  Jan;48(4):757--773.

\bibitem{gaube_satellite_2013}
Gaube P, Chelton DB, Strutton PG, Behrenfeld MJ.
\newblock Satellite observations of chlorophyll, phytoplankton biomass, and
  {Ekman} pumping in nonlinear mesoscale eddies.
\newblock Journal of Geophysical Research: Oceans. 2013;118(12):6349--6370.

\bibitem{chakraborty_harmful_2014}
Chakraborty S, Feudel U.
\newblock Harmful algal blooms: combining excitability and competition.
\newblock Theoretical Ecology. 2014 Aug;7(3):221--237.

\bibitem{jung_application_1993}
Jung C, Tel T, Ziemniak E.
\newblock Application of scattering chaos to particle transport in a
  hydrodynamical flow.
\newblock Chaos: An Interdisciplinary Journal of Nonlinear Science. 1993
  Oct;3(4):555--568.

\bibitem{martin_patchy_2002}
Martin AP, Richards KJ, Bracco A, Provenzale A.
\newblock Patchy productivity in the open ocean.
\newblock Global Biogeochemical Cycles. 2002;16(2):9--1--9--9.

\bibitem{barton_canary_2001}
Barton ED.
\newblock Canary {And} {Portugal} {Currents}.
\newblock In: Steele JH, editor. Encyclopedia of {Ocean} {Sciences}. Oxford:
  Academic Press; 2001. p. 380--389.

\bibitem{platt_-off_1993}
Platt N, Spiegel EA, Tresser C.
\newblock On-off intermittency: {A} mechanism for bursting.
\newblock Physical Review Letters. 1993 Jan;70(3):279--282.

\bibitem{steele_simple_1981}
Steele JH, Henderson EW.
\newblock A {Simple} {Plankton} {Model}.
\newblock The American Naturalist. 1981 May;117(5):676--691.

\bibitem{edwards_zooplankton_1999}
Edwards AM, Brindley J.
\newblock Zooplankton mortality and the dynamical behaviour of plankton
  population models.
\newblock Bulletin of Mathematical Biology. 1999 Mar;61(2):303--339.

\bibitem{edwards_adding_2001}
Edwards AM.
\newblock Adding {Detritus} to a {Nutrient}–{Phytoplankton}–{Zooplankton}
  {Model}:{A} {Dynamical}-{Systems} {Approach}.
\newblock Journal of Plankton Research. 2001 Apr;23(4):389--413.

\end{thebibliography}

\newpage
\appendix

\section*{Supplemental Material}
\subsection*{Hydrodynamic model}
The analytically defined model describes the velocity field for an
incompressible viscid fluid with a Reynolds number at which the solution of the
Navier-Stokes equation is time periodic. The period of the flow is $T$. During
this time, two vortices are created in the wake, with a phase shift of $T/2$,
and move away from the island. The two vortices rotate in opposite directions
and are characterized by a vortex strength $\omega$. One of them travels
slightly above and the other slightly below the axis at $y_0$. Please note that
the assumption of a two dimensional velocity field relies on the fact that the
vertical velocities in the ocean are significantly smaller compared to the
horizontal ones. Additional dynamical properties of the flow relevant to this
work are reviewed in Sec. 3.1.

According to the situation in this geographical region the period of the flow
$T$ is $32$ days. The parameters used for the flow field are shown in
Table.~\ref{tab:hyd}. Also following ~\cite{sandulescu_kinematic_2006} we
superimpose the Ekman flow $u_E$ in the $y$ direction, perpendicular to the main
flow, for $x > 1$, see Fig.1 (a) of the main text.

\begin{table}[h!]
  \begin{center}
 \begin{tabular}{|p{20em} c c|} 
   \hline
   Parameter & Symbol & Used value \\[0.5ex]
   \hline\hline
   Island radius & $r$ & $25$ km \\ [1.5ex] 
   Horizontal main flow velocity & $u_0$ & $0.18$ m s$^{-1}$  \\[1.5ex] 
   Velocity of the Ekman flow & $u_E$ & $0.018$ m s$^{-1}$\\ [1.5ex] 
   Vortex strength  & $\omega$ & $55 \cdot 10^3$ km$^2$ s$^{-1}$  \\[1.5ex] 
 \hline
 \end{tabular}\caption{Parameters used in the hydrodynamic flow model (for details of parameters see ~\cite{sandulescu_kinematic_2006}).}\label{tab:hyd}
\end{center}
\end{table}

\subsection*{Biological model}
The functional responses used and the parameters are listed in
Table~\ref{tab:func} and Table~\ref{tab:param} respectively. Please note that
these differential equations are based on some traditional NPZ models, such as
of Steele \& Henderson~\cite{steele_simple_1981} and Edwards \&
Brindley~\cite{edwards_zooplankton_1999}. An important characteristics of these
models is that the nutrient uptake by phytoplankton $f_{N, T}(N)$ ($f_N(N)$ for
the non-toxic species and $f_T(N)$ for the toxic species) is given by a Holling
Type II functional response, while the grazing of zooplankton $h$ considers a
Holling Type III functional response (see Table~\ref{tab:func}). Additional
effects of interspecific and intraspecific competition are given by the function
$g$, where $a/b$ is the maximum nutrient uptake rate of phytoplankton averaged
over the depth of the mixed layer.  The differences between the two groups of
phytoplankton can be introduced through different parameters: their nutrient
conversion rates $\theta_{N, T}$, half saturation constants $e_{N, T}$,
respiration rates $r_{N, T}$, their feeding preference by zooplankton, $\phi$,
and their quality as food for zooplankton expressed by the conversion rates
$\alpha_{N, T}$. However, please note, that there is no direct influence of the
toxic species on the mortality of zooplankton. Therefore this model is not
restricted to HAB formation, but can be also used to describe the emergence of
any phytoplankton bloom, in which the two different competing species are
involved. The notation of toxic and non-toxic species simplifies the extension
of our findings. The recycling by bacteria is considered indirectly with factors
$\beta$ and $\gamma$ for conversion of the dead material back into nutrients.

\begin{table}
  \begin{center}
 \begin{tabular}{|p{20em} c c|} 
   \hline
   & & \\  
   Nutrient uptake  by non-toxic species of phytoplankton  & $f_N(N)$ & $\frac{N}{e_N + N}$ \\ [1.5ex]
   Nutrient uptake  by toxic species of phytoplankton  & $f_T(N)$ & $\frac{N}{e_T + N}$ \\ [1.5ex] 
   Growth rate limitation due to light attenuation& $g(P_N, P_T)$ & $\frac{a}{b+cP_N + cP_T}$  \\[1.5ex] 
   Feeding rates of the Zooplankton & $h(P_N, P_T)$ & $\frac{1}{(\mu^2 +P_N^2 + P_T^2)}$ \\ [1.5ex] 
 \hline
 \end{tabular}\caption{Functional responses used in Eq.(1) of the main text}\label{tab:func}
\end{center}
\end{table}

 \begin{center}
\begin{table}[h!]

 \begin{tabular}{|p{18em} p{2.5em} p{5em} p{5em} p{4em} |} 
 \hline
 Parameter & Symb. & Sys.(I) & Sys.(II) & Units \\[0.5ex] 
 \hline\hline
 a/b maximum daily nutrient uptake & $a$ & $0.2$ & $0.2$ &m$^{-1}$day$^{-1}$\\ 

   Light attenuation by water & $b$ & $0.1$& $0.1$ & m$^{-1}$  \\

   Phytoplankton self-shading coefficient & $c$ & $0.4$& $0.4$& m$^{2}$ gC$^{-1}$\\

   Mortality rate of Zooplankton & $d$ & $0.065$&  $0.065$& day$^{-1}$\\
   
   Half-saturation rate for $N$ uptake of $P_N$ & $e_N$ & $0.02$& $0.02$ & gCm$^{-3}$ \\
   Half-saturation rate for $N$ uptake of $P_T$ & $e_T$ & $0.1$& $0.1$ & gCm$^{-3}$ \\
  
   Respiration rate of $P_N$ & $r_N$ & $0.1$ & $0.1$&day$^{-3}$\\
   Respiration rate of $P_T$& $r_T$ & $0.05$ & $0.1$ &day$^{-3}$\\
   Conversion rate of nutrients into $P_N$ & $\theta_N$ & $0.4$ & $0.4$ &\\
   Conversion rate of nutrients into $P_T$ & $\theta_T$ & $0.8$ & $0.4$ &\\
   Phytoplankton sinking rate & $s$ & $0.08$ & $0.08$ &day$^{-1}$ \\
   Growth efficiency of $Z$ due to $P_N$ & $\alpha_N$ & $0.25$ & $0.5$& \\
   Growth efficiency of $Z$ due to $P_T$ &  $\alpha_T$ & $0.2$ & $0.2$&\\
   $Z$ excretion fraction & $\beta$ & $0.33$ & $0.33$ &\\
   Excretion factor of $Z$ & $\gamma$ & $0.5$ & $0.5$&\\
   Maximum grazing rate of $Z$ & $\lambda$ & $0.65$ & $1.3$ &day$^{-1}$ \\
   Grazing of $Z$ half saturation constant & $\mu$ & $0.02$ & $0.02$ & gC m$^{-3}$ \\
   Intensity of grazing on $P_T$ & $\phi$ & $0.5$ & $0.05$ &\\[1ex] 
   \hline
 \end{tabular}\caption{The values used are taken from ranges given in~\cite {edwards_adding_2001}}\label{tab:param}
 \end{table}
\end{center}

\subsection*{Coupled Model}

The full system of equations is solved using a semi-Lagrangian
algorithm~\cite{sandulescu_biological_2008}, the code can be found at
https://github.com/kseniaguseva/Upwelling.  We use a grid of $[500 \times 300]$
points, the integration is carried out for step size $dt = 0.01$, and the
diffusion step $dt_D = \frac{dt}{10}$ which guaranties that
$\frac{D dt}{dx dy} < 0.5$, so that the stability condition of the integrator is
fulfilled.

Everywhere in the observation area except for the prescribed upwelling region
located above the island, the cross thermocline exchange rate is set to
$k_d$. In the upwelling region the value is exchanged between $k_d$ and
$k_{\text{up}}$ in time, corresponding to intermittent upwelling events. The
upwelling region, if not stated otherwise, spans the region: $x \in [-1, 1]$ and
$y \in [2, 2.5]$, see Fig.1 of the main text.

All the modeled species enter the system from the left at $x = -2$ with the same
concentration at all $y$ values and are advected across the observation area. We
assume that they arrive from the open ocean, an environment poor in nutrients
and plankton. Therefore we use as the influx $20 \%$ of the steady state
concentration value reached by each given species for a cross thermocline
exchange rate $k_d$. Please note that all species of phytoplankton are present
in the system in the influx. As we show in Sec.2.2 of the main text the
non-toxic species dominates in the influx conditions, since those are low in
nutrients, for both scenarios that we analyse.

\section*{Biological model with hydrodynamics, in the absence of upwelling}~\label{sec:noupw}

Hare we present the dynamics of the coupled biological-hydrodynamic model
without upwelling to allow a comparison with the results in the main text.
First, we can mention that system (I) has very similar properties as the
population dynamics analysed in \cite{bastine_inhomogeneous_2010}.  But in
contrast to \cite{bastine_inhomogeneous_2010} we are here more interested in the
dynamics of plankton in the whole area and not only in the development of a
plankton bloom related to specific regions such as vortices. When the upwelling
is not present the non-toxic species is found in higher concentrations in a
region around the island and inside every vortex formed in the wake. This
spatial distribution of the non-toxic species reflects the nutrient accumulation
in those regions, which results from the vertical exchange of nutrients across
the thermocline while the horizontal advection is slow. Subsequently, this high
nutrient concentration is captured by the vortices behind the island, where it
creates good conditions for the growth of the non-toxic species. Nevertheless
this accumulation of nutrients does not reach the threshold to trigger a
HAB. Notably, system (I) is characterized by a periodic timeseries for the
spatial average of the non-toxic species following the vortex formation and
advection: $\left<P_N\right> $ has a period $T/2$.  For system (I) the averages
correspond to: $\left<P_N\right>^* = 0.042$ gC m$^{-3}$,
$\left<P_T\right>^* = 0.019$ gC m$^{-3}$ and $\left<Z\right>^* = 0.005$ gC
m$^{-3}$, $\left<N\right>^* = 0.025$ gC m$^{-3}$.

We continue with the characterization of the spatio-temporal patterns formed in
system (II). For the scenario without upwelling the non-toxic species, as in the
previous case, develops around the island and inside the vortices. As in system
(I), the growth of phytoplankton also follows the large concentration of
nutrients. The large presence of non-toxic species creates now good conditions
for the development of zooplankton, that also grows in the same location but at
a slower rate. As a result, this system displays more complex spatio-temporal
patterns, where as the vortices are advected by the flow field the concentration
of zooplankton increases as well. On the other hand due to the large
phytoplankton growth these vortices become depleted in nutrients. Therefore,
without upwelling, conditions that would give an advantage to the development of
the toxic species are never met. In summary, in system (II) the non-toxic
species is the dominant species everywhere in space with the average
concentration $\left<P_N\right>^* = 0.0084$ gC m$^{-3}$, which is in agreement
with low nutrient concentration in the system $\left<N\right>^* = 0.0286$. The
average concentration of the other plankton species are:
$\left<P_T\right>^* = 0.0010$ gC m$^{-3}$ and $\left<Z\right>^* = 0.0054$ gC
m$^{-3}$. Furthermore, the average concentration of all species is about one
order of magnitude smaller than for system (I). This is a direct consequence of
the used values of the thermocline exchange rate.

\section*{Nutrient release from the upwelling region}

\begin{figure}[h!]
  \centering
  \includegraphics[scale=1.]{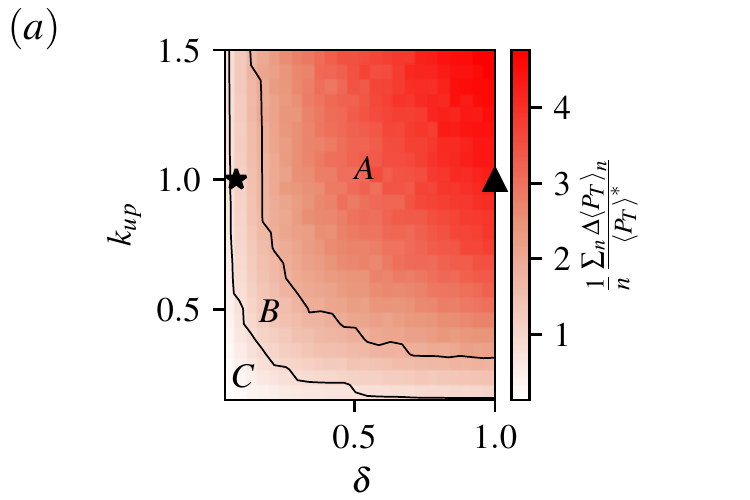}
  \includegraphics[scale=1.]{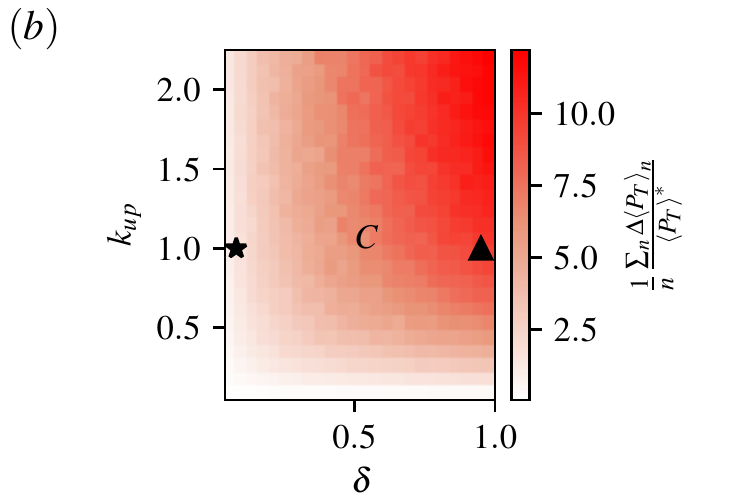}
  \caption{Responses of the toxic species to upwelling events of different
    duration $\delta$ and $k_{\text{up}}$ in: (a) system(I) and (b) in system
    (II).  The region A corresponds to $P_{\text{HAB}} > 0.98$, B to
    $0.1< P_{\text{HAB}} <0.98$ and C to $P_{\text{HAB}} < 0.1$. The values were
    computed from a time series with $150$ upwelling events distributed over
    $600$ T. The stars mark $k_{\text{up}} = 1.0$ and $\delta = 0.08$, the
    histograms of the responses of the two phytoplankton species to upwelling at
    this point are shown in Fig.10 (a,b) of the main text. The triangles mark
    $k_{\text{up}} = 1.0$ and $\delta = 0.95$, for the respective histograms see
    Fig.10 (c,d) of the main text.}\label{fig:puls1_total}
\end{figure}

Finally, we investigate the last parameter that controls the release of
nutrients from the upwelling region --- the cross thermocline exchange rate
$k_{\text{up}}$. To get some insight of the effect of $k_{\text{up}}$ on the
growth of toxic species we compare the $\Delta \left<P_T\right>_n$ to the
undisturbed concentration value $\left<P_T\right>^*$. More precisely we
calculate
$\frac{1}{n}\sum_n\frac{\Delta \left<P_T\right>_n}{\left<P_T\right>^*}$, for
$n = 150$ upwelling events for a range of $k_{\text{up}}$ and $\delta$ values.
As expected small values of either $k_{\text{up}}$ or $\delta$ do not trigger
the growth of the toxic species. By contrast, for high values of these
parameters, the toxic species exhibits high growth in both
systems. Fig.~\ref{fig:puls1_total} shows the impact of these two parameters
simultaneously. Please note that $k_{\text{up}}$ may limit the HAB formation
given a fixed $\delta$. The mechanism behind that is simple: the smaller
$k_{\text{up}}$, the more time is needed for the released nutrients to reach a
significant level for the toxic species development. This means that for a small
value of $k_{\text{up}}$ the initial times of upwelling events are even more
constrained, i.e.  they have to start at the beginning of vortex formation to
allow for sufficient input of nutrients. In Fig.~\ref{fig:puls1_total} we also
show how the dominance change of species depends on parameters $\delta$ and
$k_{\text{up}}$. To this end we compute the probability, $P_{\text{HAB}}$, that
the toxic species out-competes the non-toxic one as a result of an upwelling
event,
$P_{\text{HAB}}= P(\text{max}(\left<P_T\right>)_n > \text{max}(\left<P_N
\right>_n) )$. This probability again is computed using the time series with
$200$ upwelling events. In Fig.~\ref{fig:puls1_total} (a) we delineate the
region $A$, where $P_{\text{HAB}}> 0.98$. In this region most of the upwelling
events lead to HABs. The region $C$ is restricted to parameters where
$P_{\text{HAB}} <10 \%$, while the intermediate region $B$ spreads over the
parameter range where many different outcomes are possible. Note that the sizes
of these regions depend on the parameters of the biological model. Since in the
system (I) mainly the toxic species responds to the upwelling event, the largest
part of the parameter spaces is covered by the region A. By contrast, as we have
explained before, the dominance change in system (II) is restricted to some
spatial regions and therefore, region $C$ spans over the whole parameter space
for this scenario, see Fig.~\ref{fig:puls1_total} (b).

\section{Intermittency}
To model the time series of intermittent upwelling events we use the absolute
value of the $x_1$ variable from the system of the following differential
equations:

\begin{equation}\label{eq:platt} 
  \begin{split}
  \dot{x_1}& = x_2,\\
  \dot{x_2}& = -x_1^3 - 2 x_1 x_3 + x_1x_5 - \mu x_2,\\
  \dot{x_3}& = x_4,\\
  \dot{x_4}& = -x_3^3 - \nu_1x_1^2 + x_3x_3 - \nu_2x_4,\\
  \dot{x_5} &= - \nu_3x_5 - \nu_4x_1^2 - \nu_5(x_3^2 - 1),\\
  \end{split}
\end{equation}

where $\mu = 1.815$, $\nu_1 = 1.$, $\nu_2= 1.815$, $\nu_3 = 0.44$,
$\nu_4 = 2.86$, $\nu_5 =2.86$. For these parameters the system displays on-off
intermittency, for details see~\cite{platt_-off_1993}.

To lead to upwelling events of the adequate duration we rescale the time in
Eqs.(\ref{eq:platt}) using $t' = t/T$, where T is the period of the flow
field. The ``off'' states of the time series from Eqs.(\ref{eq:platt}) are
characterized by $|x_1| \sim 0$, and the ``on'' state by max$(|x_1|) \sim
1$. Therefore, we transform this time series to:
\begin{equation}
  k(t') = (k_{\text{up}} - k_d) |x_1(t')|  + k_d.
\end{equation}

With this transformation the new time series has the `` off`` state $\sim k_d$
and an ``on'' state which can be at most $k_{\text{up}}$.  The time series
$k(t')$ is used to define the cross thermocline exchange rate at the upwelling
region.

\end{document}